\documentclass[final,leqno,onefignum,onetabnum]{siamltex1213}

\newcommand{\MyTitel}{MASSIVELY PARALLEL ALGORITHMS FOR THE LATTICE BOLTZMANN METHOD ON NON-UNIFORM GRIDS}

\newcommand{\MyAbbreviatedTitel}{PARALLEL LATTICE BOLTZMANN ON NON-UNIFORM GRIDS}
\newcommand{\MyKeywords}{lattice Boltzmann method, grid refinement, non-uniform grids, supercomputing, scalable parallel algorithms, parallel performance, LBM, HPC, CFD}

\usepackage[numbers,sort]{natbib}
\usepackage{amssymb}
\usepackage{amsmath}
\usepackage{xfrac}
\usepackage{siunitx}

\usepackage[utf8]{inputenc}
\newcommand{\Walberla}{\textsc{waLBerla}}
\newcommand{\MLUPSPS}{\mbox{MLUPS/$sec$}}
\newcommand{\MLUPSPSC}{\mbox{MLUPS\hspace{.08em}/\hspace{.08em}$sec\!\cdot\!core$}}

\usepackage[ruled,lined,longend,linesnumbered]{algo2e}
\usepackage{booktabs}
\usepackage[normalem]{ulem}

\hypersetup{
    pdfstartview={Fit},             
    pdftitle={\MyTitel{}},          
    pdfauthor={Florian Schornbaum}, 
    pdfsubject={\MyTitel{}},        
    pdfkeywords={\MyKeywords{}},    
    hidelinks
}

\title{\MyTitel{}}

\author{Florian Schornbaum\footnotemark[2] \and Ulrich R\"ude\footnotemark[2]}

\begin{document}

\maketitle

\renewcommand{\thefootnote}{\fnsymbol{footnote}}
\footnotetext[2]{Chair for System Simulation, Friedrich-Alexander-Universit\"at Erlangen-N\"urnberg, Erlangen, Germany
(\email{florian.schornbaum@fau.de, ulrich.ruede@fau.de}).}
\renewcommand{\thefootnote}{\arabic{footnote}}

\slugger{sisc}{xxxx}{xx}{x}{x--x}

\begin{abstract}
The lattice Boltzmann method exhibits excellent scalability on
current supercomputing systems and has thus
increasingly become an alternative method for large-scale non-stationary
flow simulations, reaching up to a trillion ($10^{12}$) grid nodes.
Additionally, grid refinement can lead to substantial savings in memory and compute time.
These saving, however, come at the cost of much more complex data structures and algorithms.
In particular, the interface between subdomains with different grid sizes must receive special treatment.
In this article, we present parallel algorithms, distributed data structures, and communication routines that
are implemented in
the software framework \Walberla{}
in order to support large-scale, massively parallel lattice Boltzmann-based simulations on non-uniform grids.
Additionally, we evaluate the performance of our approach on two current petascale supercomputers.
On an IBM Blue Gene/Q system,
the largest weak scaling benchmarks with refined grids are executed with almost two million threads,
demonstrating not only near-perfect scalability but also an
absolute performance of close to a trillion lattice Boltzmann cell updates per second.
On an Intel-based system, the strong scaling of a simulation with refined grids and a total of
more than 8.5 million cells is demonstrated to reach a performance of less than one millisecond per time step.
This enables simulations with complex, non-uniform grids and four million time steps per hour compute time.
\end{abstract}

\begin{keywords}
\MyKeywords{}
\end{keywords}

\begin{AMS}
65C, 
65Y, 
68W  
\end{AMS}

\pagestyle{myheadings}
\thispagestyle{plain}
\markboth{FLORIAN SCHORNBAUM AND ULRICH R\"UDE}{\MyAbbreviatedTitel}

\section{Introduction}

In the last decade, the lattice Boltzmann method (LBM) has gained popularity
as an alternative to classical Navier-Stokes solvers for computational fluid dynamics
(CFD)~\cite{Chen98,Aidun2010}.
For simulations with the LBM, the simulation domain 
is typically discretized with a uniform Cartesian grid.
If the resolution of a three-dimensional simulation must be increased
in space and time, then the total number of cells
and the computational cost increase quickly.

Many of the existing frameworks for the LBM \cite{Feichtinger2011105,Hasert2014784,palabos,heuveline2007,lb3d,Groen2013412,Randles2013,Desplat2001273}
are therefore designed for parallel computers
where many show excellent scalability, i.e., their performance increases linearly
with the number of processors employed.
Going beyond just scalability, a carefully crafted architecture-aware implementation
of the LBM as realized in the \Walberla{} framework~\cite{Godenschwager2013} can achieve excellent absolute performance and thus reduce the time to solution
and the power consumption to reach a given computational objective.
Using globally uniform Cartesian grids, i.e., using the same grid resolution for the entire simulation,
it is possible to discretize flow geometries with in excess of $10^{12}$ lattice cells on current petascale supercomputers~\cite{Godenschwager2013}.

However, for certain simulations, only parts of the entire domain require high resolution.
In order to focus the computational resources on those regions that need
high accuracy, many advanced methods in CFD rely on grid refinement.
For incorporating grid refinement into the LBM, node-based~\cite{Filippova1998219,Kandhai2000100,Lagrava20124808,FLD:FLD280,Peng2006460,Toelke2006820}
and volume-based~\cite{Rohde2006,Chen2006,Neumann2013,Hasert2014784,Guzik2014} approaches have been proposed.

In node-based approaches, the distribution functions of the LBM are located at the nodes of the lattice cells.
At the interface between two different grid resolutions, fine grid nodes either coincide with or are located exactly halfway between coarse nodes.
For the LBM scheme of \cite{Toelke2006820}, a parallel extension was proposed in \cite{Freudiger08}.
In node-based approaches,
the non-equilibrium part of the distribution functions typically must be rescaled at the interface between different grid resolutions.

In volume-based approaches, the distribution functions of the LBM are located at the center of lattice cells.
During refinement, coarse cells are uniformly subdivided into multiple finer cell.
As a result, the centers of fine and coarse cells will not coincide.
Volume-based grid refinement approaches for the LBM allow formulations that
guarantee the conservation of mass and momentum without
the need to rescale the non-equilibrium part of the distribution functions.
For volume-based grid refinement, \cite{Neumann2013} and \cite{Hasert2014784} presented parallelization
approaches that are suitable for large-scale systems.
In order to achieve scalability to these systems, both approaches are based on a tree partitioning of the simulation space.
\cite{Hasert2014784} introduces the MUSUBI software that relies on a linearized octree,
\cite{Neumann2013} uses the Peano framework~\cite{Bungartz2010} that is based on a more generalized spacetree concept.
An octree-based domain partitioning approach was also proposed by~\cite{fietz2012}, however,
not for incorporating grid refinement but for better fitting block-based data structures to complex geometries.
To our best knowledge, other popular simulation codes based on the LBM such as Palabos~\cite{palabos,Lagrava20124808}, OpenLB~\cite{heuveline2007,heuveline2009,krause2010}, LB3D~\cite{lb3d,groen2011},
HemeLB~\cite{Groen2013412}, HARVEY~\cite{Randles2013}, or LUDWIG~\cite{Desplat2001273} are at a state that they either do not support grid refinement,
only support grid refinement for 2D problems, or have not yet demonstrated large-scale simulations on non-uniform grids.

In this article, we present a volume-based refinement approach that consists of a distributed memory parallelization
of the algorithm described in \cite{Rohde2006} combined with the interpolation scheme suggested by \cite{Chen2006}.
The main goal of the implementation into the software framework \Walberla{} is to ensure
applicability for current petascale and future exascale supercomputers.
To this end, all algorithms and data structures are carefully designed
with high node level efficiency as well as scalability to massively parallel systems in mind.
We demonstrate simulations on non-uniform grids with multiple different grid
resolutions and close to one trillion ($10^{12}$) cells that run on current petascale systems
and make use of up to nearly two million concurrent threads.
We also show that our implementation can reach throughput rates of more than
1,000 time steps per second in a strong scaling scenario,
enabling simulations that perform several million time steps per hour compute time.
To the best knowledge of the authors, the total number of cells that can be handled
as well as the overall performance achieved
significantly exceed the data
published for the LBM on non-uniform grids to date \cite{Freudiger08,Schoenherr20113730,FLD:FLD2469,Hasert2014784}.
In addition to the algorithms and data structures,
we also propose a method for scaling the two relaxation parameters of the two-relaxation-time collision model across different grid resolutions.

The remainder of this article is organized as follows.
In section~\ref{sec:LBM:general}, we give a brief introduction to the LBM including details about all the necessary extensions that are required for incorporating grid refinement.
In section~\ref{sec:pconcept}, we describe
new data structures and concepts that we added to the \Walberla{} framework in order to extend the framework to
support efficient, massively parallel simulations on non-uniform grids.
Details on the implementation of our parallel algorithm for the LBM, communication, and load balancing follow in section~\ref{sec:lbmref}.
In section~\ref{sec:benchmarks}, we verify the correctness of our parallel scheme for two different Poiseuille flow scenarios and
present weak and strong scaling benchmarks that demonstrate the performance of our approach on two petascale supercomputers.
We conclude the article in section~\ref{sec:conclusion}.

\section{The lattice Boltzmann method}\label{sec:LBM:general}

\begin{figure}[tbp]
  \centering
  \includegraphics[width=0.4\textwidth]{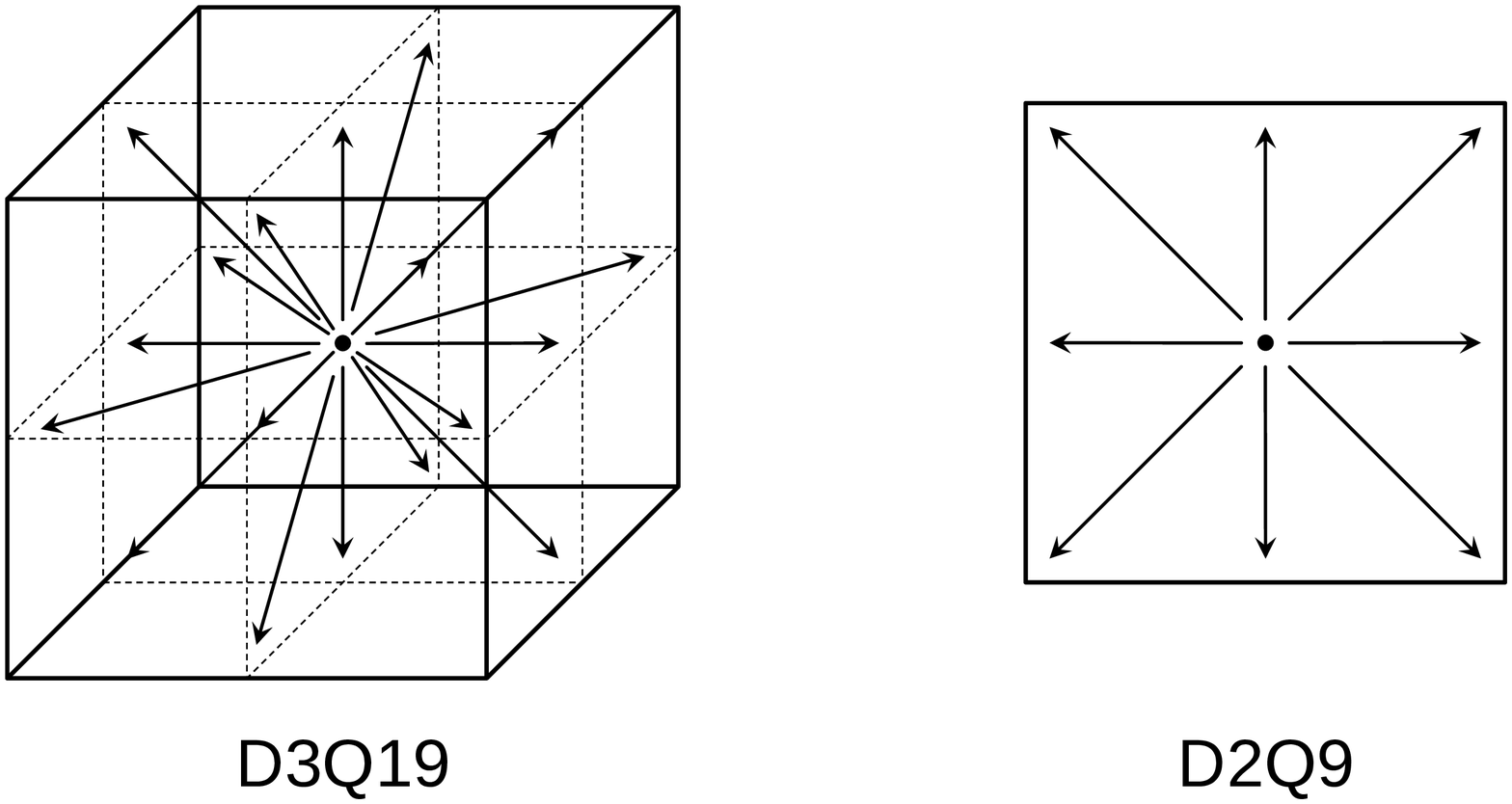}
  \caption{Schematic visualization of one lattice cell for the three-dimensional D3Q19 and the two-dimensional D2Q9 model.
For D3Q19, no data is exchanged across cell corners, but only across edges and faces.
While all simulations of this article use the D3Q19 model,
all later illustrations are in 2D in order to facilitate the schematic visualization of the methods.}
  \label{fig:dxqx}
\end{figure}

The LBM uses an explicit time stepping scheme that is well suited for extensive parallelization due to its data locality.
In each time step, information is only exchanged between neighboring grid cells.
Each cell stores $N$ particle distribution functions (PDFs)
\begin{equation*}
f_\alpha: \Omega \times T \mapsto [0,1],~~~\alpha = 0,...,N-1,
\end{equation*}
where $\Omega \subset \mathbb{R}^3$ and $T = [0,t_{end}]$ are the physical and time domain, respectively.
In this article, we employ the three-dimensional D3Q19 model as it was originally developed by
Qian, d`Humi\'eres, and Lallemand~\cite{Qian1992} with $N=19$ PDFs stored in each grid cell (see Figure~\ref{fig:dxqx}).
In general, the lattice Boltzmann equation can then be written as
\begin{equation*}
f_\alpha(\mathbf{x}_i+\mathbf{e}_\alpha \Delta t,t+\Delta t) - f_\alpha(\mathbf{x}_i,t) = \mathcal{C}_\alpha(\mathbf{f}),
\end{equation*}
with $\mathbf{x}_i$ denoting the center of the i-th cell in the discretized simulation domain,
$\mathbf{e}_{\alpha}$ denoting the discrete velocity set $\{\mathbf{e}_{\alpha}|\alpha = 0,\ldots,N-1\}$, $t$ denoting the current time step,
$\Delta t$ denoting the time step size, and $\mathcal{C}_\alpha(\mathbf{f})$ denoting the collision operator of the LBM.
Algorithmically, the LB equation is typically separated into a collision~(\ref{eq:LBM:collide}) and a streaming step~(\ref{eq:LBM:stream})
\begin{subequations}
\begin{align}
\tilde{f_\alpha}(\mathbf{x}_i,t) &= f_\alpha(\mathbf{x}_i,t) + \mathcal{C}_\alpha(\mathbf{f}) \label{eq:LBM:collide},\\
f_\alpha(\mathbf{x}_i+\mathbf{e}_\alpha \Delta t,t+\Delta t) &= \tilde{f_\alpha}(\mathbf{x}_i,t) \label{eq:LBM:stream},
\end{align}
\end{subequations}
with $\tilde{f_\alpha}$ denoting the post-collision state of the distribution function.
During streaming, PDFs are exchanged only between neighboring cells while the collision step is a local operation in each cell.
The three most commonly used collision schemes are the single-relaxation-time (SRT/LBGK) model~\cite{Bhatnagar1954},
the two-relaxation-time (TRT) model~\cite{Ginzburg2008, Ginzburg2008a},
and the multiple-relaxation-time (MRT) model~\cite{d1994generalized, humieres2002}.
In this article, we focus on the SRT and the TRT model.
For the SRT model, the collision operator is given by
\begin{equation*}
\mathcal{C}_\alpha(\mathbf{f}) = -\frac{\Delta t}{\tau}(f_\alpha(\mathbf{x}_i,t)-f_\alpha^{eq}(\mathbf{x}_i,t)) = - \omega (f_\alpha(\mathbf{x}_i,t)-f_\alpha^{eq}(\mathbf{x}_i,t)),
\end{equation*}
where $\tau$ is the relaxation time, $\omega \in ]0,2[$ the dimensionless relaxation parameter,
and $f_\alpha^{eq}(x_i,t)$ is the equilibrium distribution.
The relaxation time $\tau$ and the relaxation parameter $\omega$ are related to the kinematic viscosity $\nu$ with
\begin{equation}
\nu = c_s^2 \left( \tau - \frac{\Delta t}{2} \right) = \frac{c^2}{3} \left( \frac{1}{\omega} - \frac{1}{2} \right) \Delta t \label{eq:LBM:viscosity_srt},
\end{equation}
where $c=\Delta x / \Delta t$ is the lattice velocity and $c_s=c / \sqrt{3}$ is the speed of sound for an isothermal fluid~\cite{Chen98, Aidun2010}.
For the incompressible LBM~\cite{He97}, the equilibrium distribution function can be calculated according to
\begin{equation}
f_\alpha^{eq}(\mathbf{x}_i,t) = w_\alpha \left[ \rho(\mathbf{x}_i,t) + \rho_0 \left( \frac{3 \mathbf{e}_{\alpha} \cdot \mathbf{u}(\mathbf{x}_i,t)}{c^2} + \frac{9 ( \mathbf{e}_{\alpha} \cdot \mathbf{u}(\mathbf{x}_i,t) )^2}{2 c^4} - \frac{3 \mathbf{u}(\mathbf{x}_i,t)^2}{2 c^2} \right) \right], \label{eq:LBM:fequ}
\end{equation}
with $w_\alpha$ denoting the lattice model-specific weighting factor corresponding to $\mathbf{e}_{\alpha}$ and $\rho_0$ denoting the reference density.
$\rho(\mathbf{x}_i,t)$ and $\mathbf{u}(x_i,t)$ are the fluid density and the fluid velocity, respectively. They are calculated from the first two moments
\begin{equation}
\rho(\mathbf{x}_i,t) = \sum_{\alpha} f_\alpha^{eq}(\mathbf{x}_i,t) \text{~~and~~} \mathbf{u}(\mathbf{x}_i,t) = \frac{1}{\rho_0} \sum_{\alpha} \mathbf{e}_{\alpha} f_\alpha^{eq}(\mathbf{x}_i,t) \label{eq:LBM:fvelocity} .
\end{equation}
For the TRT model, the distribution functions are split into
a symmetric (even) and an asymmetric (odd) part
\begin{equation*}
f_\alpha^{\pm} = \frac{1}{2}(f_\alpha \pm f_{\bar{\alpha}})~,~f_\alpha^{eq\pm} = \frac{1}{2}(f_\alpha^{eq} \pm f_{\bar{\alpha}}^{eq}),
\end{equation*}
with $\bar{\alpha}$ denoting the inverse direction of $\alpha$.
The corresponding collision operator is given by
\begin{equation*}
\mathcal{C}_\alpha(\mathbf{f}) = -\lambda_e(f_\alpha^{+}-f_\alpha^{eq+}) - \lambda_o(f_\alpha^{-}-f_\alpha^{eq-}),
\end{equation*}
with $\lambda_e \in ]0,2[$ and $\lambda_o \in ]0,2[$ denoting the even and the odd relaxation parameter of the TRT model. If
$\lambda_e = \lambda_o = \omega = \Delta t / \tau$,
the TRT model coincides with the SRT model.
The kinematic viscosity is related to $\lambda_e$ just like it is related to $\omega$ when using the SRT model. Hence, for the TRT model,
\begin{equation}
\nu = c_s^2 \left( \frac{1}{\lambda_e} - \frac{1}{2} \right) \Delta t \label{eq:LBM:viscosity_trt}.
\end{equation}
Typically, implementations of the LBM are formulated in a dimensionless lattice space.
In order to set up simulations and evaluate the results,
physical quantities must then be transformed from physical space into dimensionless lattice units, and vice versa~\cite{lattlbm2008}.

\subsection{External forces}\label{sec:LBM:forces}

External forces can be
incorporated into the LBM 
by including an additional term in
the collision step~(\ref{eq:LBM:collide})
\begin{equation*}
\tilde{f_\alpha}(\mathbf{x}_i,t) = f_\alpha(\mathbf{x}_i,t) + \mathcal{C}_\alpha(\mathbf{f}) + \Delta t F_\alpha .
\end{equation*}
There exist various different force models for the LBM~\cite{Mohamad2010990}.
Depending on the force model, the calculation of the fluid velocity~(\ref{eq:LBM:fvelocity}) and/or the velocity terms in~(\ref{eq:LBM:fequ})
must be adapted. In order to incorporate a constant body force $\mathbf{F}$ caused by a globally constant acceleration $\mathbf{a}$ into the simulation, we use
\begin{equation*}
F_\alpha = w_\alpha \frac{\mathbf{e}_{\alpha} \cdot \mathbf{F}}{c_s^2} = w_\alpha \rho_0 \frac{\mathbf{e}_{\alpha} \cdot \mathbf{a}}{c_s^2} .
\end{equation*}
For the evaluation of the equilibrium distribution~(\ref{eq:LBM:fequ}) during the collision step, we do not change the calculation of $\mathbf{u}(\mathbf{x}_i,t)$,
meaning we use an unmodified version of~(\ref{eq:LBM:fvelocity}).
However, when calculating the fluid velocity independently of the equilibrium distribution, we add an additional term to~(\ref{eq:LBM:fvelocity})~\cite{Ginzburg2008}
\begin{equation*}
\mathbf{u}(\mathbf{x}_i,t) = \frac{1}{\rho_0} \sum_{\alpha} \mathbf{e}_{\alpha} f_\alpha^{eq}(\mathbf{x}_i,t) + \frac{\Delta t \mathbf{F}}{2 \rho_0}.
\end{equation*}

\subsection{Boundary treatment}\label{sec:LBM:boundaries}

\begin{figure}[tbp]
  \centering
  \includegraphics[width=0.7\textwidth]{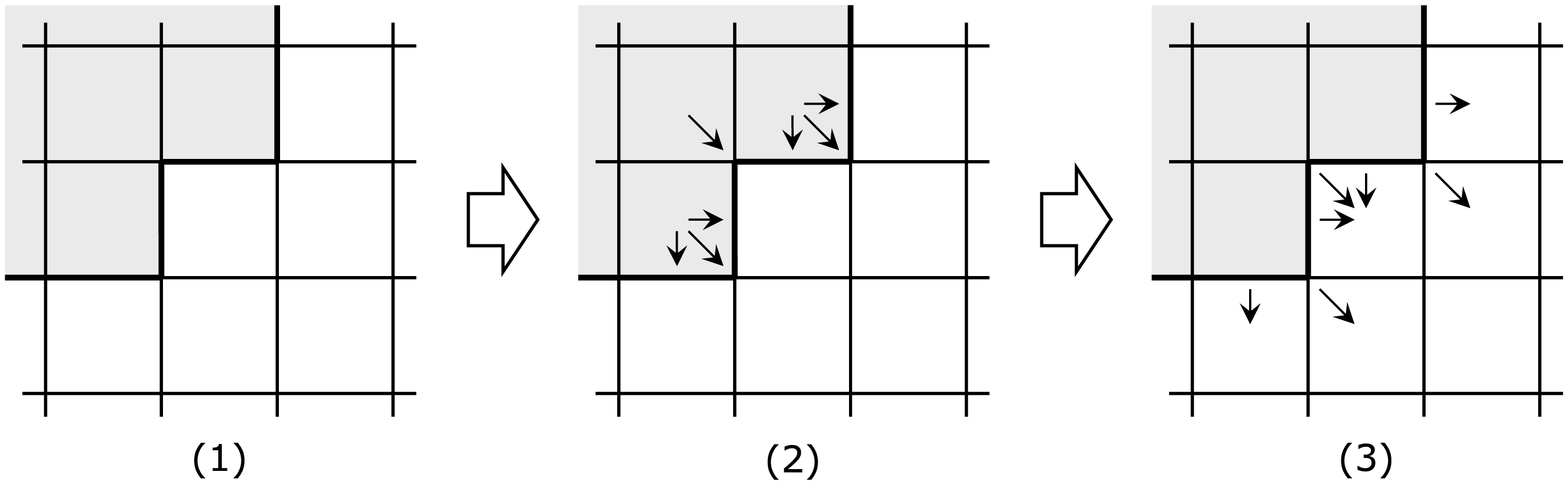}
  \caption{Boundary cells are marked in gray (1). After the collision step and prior to streaming, the boundary treatment algorithm stores PDF values inside the boundary cells itself (2). During streaming, no special treatment has to be performed for cells next to the boundary: All fluid cells can pull PDF values from all of their neighbors (3), even if the neighboring cell is a boundary cell.}
  \label{fig:boundaries}
\end{figure}

Domain boundaries and obstacles within the fluid must receive special treatment.
We mark every cell of the grid as either a fluid cell, a boundary/obstacle cell, or a cell that is outside of the simulation domain and hence does not need to be considered during the simulation at all.
Depending on which boundary marker is set for a certain cell, a different boundary treatment is performed
during a post-collision/pre-streaming step.
After the collision step, PDF values are calculated depending on the corresponding boundary conditions and are saved inside the boundary cells.
During the subsequent streaming step, these values are pulled by their neighboring fluid cells (see Figure~\ref{fig:boundaries}).

As a result of this approach, the algorithms performing the collision and the streaming step can remain unchanged
since during collision and streaming, no special operations are necessary near boundary cells.
Because of boundary treatment being a pre-streaming operation,
even existing fused stream-collide compute kernels (see section~\ref{sec:lbmref:algo}) do not require any changes.
Periodic boundaries are treated by copying PDF values from one side of the simulation domain to the other, and vice versa.
For parallel simulations, the communication step that normally synchronizes neighboring subdomains that
reside on different processes can also be used to process periodic boundaries
by establishing a communication channel between opposing sides of the simulation domain.

\subsection{Grid refinement}\label{sec:LBM:refinement}

In this article, we present an efficient, highly optimized parallelization of the algorithm proposed by~\cite{Rohde2006} for the LBM on non-uniform grids.
The algorithm presented in~\cite{Rohde2006} relies on a 2:1 balance between neighboring grid cells at the interface between two grid levels
\begin{equation*}
\Delta x_{L+1} = \frac{\Delta x_L}{2} ~~~ \Rightarrow ~~~  \Delta x_{L} = \frac{\Delta x_0}{2^L},
\end{equation*}
with $L$ denoting the grid level and $L=0$ corresponding to the coarsest level.
The jump in $L$ between a cell and all of its neighboring cells is at most 1.
A non-uniform grid conforming with these features is illustrated in Figure~\ref{fig:domain_decomposition}.
We apply acoustic scaling where the time step ratio is proportional to the grid spacing ratio \cite{Hasert:229088,Schoenherr20113730}.
Thus, due to the grid spacing ratio of 2:1, twice as many time steps must be executed on level $L+1$ compared to level $L$.
Hence, $\Delta t_L = \Delta t_0 / 2^L$.
As a result, the speed of sound $c_s=c / \sqrt{3}$ with $c=\Delta x / \Delta t$ remains constant on each grid level.
The kinematic viscosity $\nu$ (\ref{eq:LBM:viscosity_srt}) must also remain constant.
In order to keep $\nu$ constant, the dimensionless relaxation parameter $\omega$ must be properly scaled to each grid level
\begin{subequations}
\begin{align}
\nu_L = \nu_0 ~~ \Rightarrow ~~ c_s^2 \left( \frac{1}{\omega_L} - \frac{1}{2} \right) \Delta t_L &= c_s^2 \left( \frac{1}{\omega_0} - \frac{1}{2} \right) \Delta t_0 \nonumber \\
\Rightarrow \:~~ \omega_L &= \frac{2 \omega_0}{2^{L+1} + (1 - 2^L) \omega_0} \label{eq:LBM:omega_scaling_0} ,\\
\text{or more generally:~~~} \omega_L &= \frac{2^{K+1} \omega_K}{2^{L+1} + (2^K - 2^L) \omega_K} \label{eq:LBM:omega_scaling} .
\end{align}
\end{subequations}
Since for the TRT model, the kinematic viscosity is related to $\lambda_e$ just like it is related to $\omega$ when using the SRT model (\ref{eq:LBM:viscosity_trt}),
we use the same equations (\ref{eq:LBM:omega_scaling_0}) and (\ref{eq:LBM:omega_scaling}) for scaling $\lambda_e$ to different grid levels.
In order to choose $\lambda_o$, we use
\begin{equation}
\Lambda_{eo} = \left( \frac{1}{\lambda_e} - \frac{1}{2} \right) \left( \frac{1}{\lambda_o} - \frac{1}{2} \right) \label{eq:LBM:lambda_eo},
\end{equation}
with the ``magic'' parameter $\Lambda_{eo}$ suggested in~\cite{Ginzburg2008}. We propose to keep $\Lambda_{eo}$ constant on all levels and scale $\lambda_o$
on each level accordingly
\begin{equation}
\lambda_{o,L} = \frac{4 - 2 \lambda_{e,L}}{2 + (4 \Lambda_{eo} - 1)\lambda_{e,L}} \label{eq:LBM:lambda_oL} .
\end{equation}
When applying a force caused by a constant acceleration $\mathbf{a}$,
we use the parametrization of the acceleration in order to calculate level-dependent lattice acceleration $\mathbf{a}_L^*$ (in dimensionless lattice units~\cite{lattlbm2008})
\begin{equation}
\mathbf{a}_L^* = \mathbf{a} \frac{(\Delta t_L)^2}{\Delta x_L} ~~\Rightarrow~~ \mathbf{a}_L^* = \frac{\mathbf{a}_0^*}{2^L} \text{~~~or more generally:~~~} \mathbf{a}_L^* = 2^{K-L} \mathbf{a}_K^* \label{eq:LBM:force_scaling}
\end{equation}

\section{Parallelization concept and realization}\label{sec:pconcept}

In the following subsections, we give a brief introduction to the \Walberla{} framework,
present our domain partitioning concept based on a 2:1 balanced forest of octrees space decomposition,
introduce a corresponding two-stage initialization approach that proves to be useful for scalable, massively parallel simulations,
and describe the new communication patterns that are required in order to support non-uniform grid structures.

\subsection{The \Walberla{} simulation framework}\label{sec:pconcept:walberla}

At its core, \Walberla{} is a general-purpose HPC software framework that is capable of supporting different numerical methods by providing generic,
extensible concepts for domain partitioning, input and output, parallelization, and program flow control.
The main focus of the framework is on massively parallel CFD simulations based on the lattice Boltzmann method \cite{Godenschwager2013,ammer2014simulating,Bartuschat:2015:CP,C3FO60374A,Bogner201571,Pickl2014}.
It is written in C++, supports all major compilers,
and can scale from laptops and conventional desktop computers up to the largest supercomputers with millions of concurrent threads.
Besides parallelization with only OpenMP for shared memory systems or only with MPI
for distributed memory, the framework also supports hybrid parallel execution where
multiple OpenMP threads are executed per MPI process.
Fine-tuned, vectorized compute kernels combined with a performance engineering approach that
takes into account the specifics of the underlying architecture guarantee high performance~\cite{Godenschwager2013,Bauer:2015:MPP}.

\subsection{Forest of octrees domain partitioning}\label{sec:pconcept:octree}

\begin{figure}[tbp]
  \centering
  \includegraphics[width=0.7\textwidth]{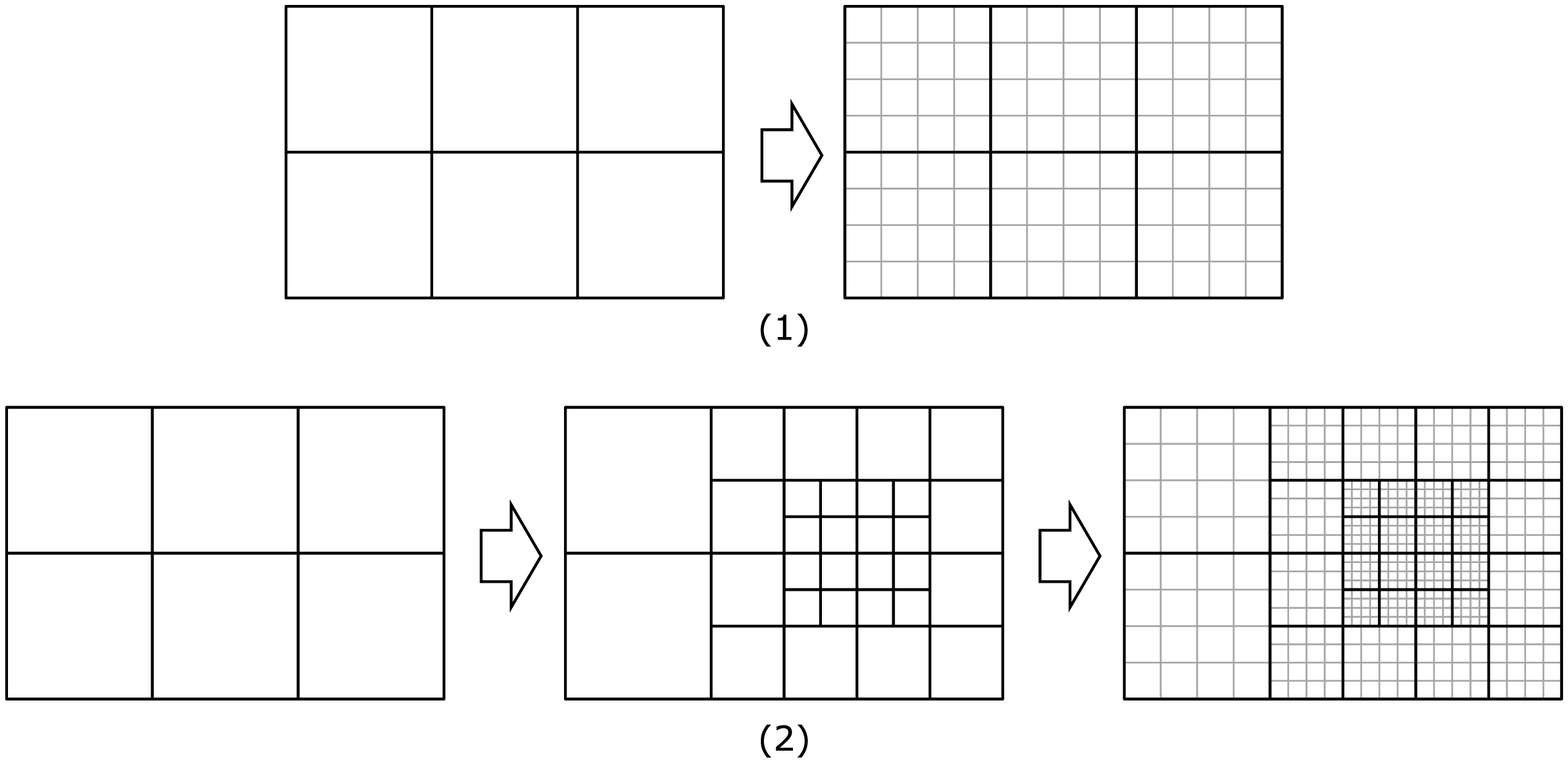}
  \caption{Without refinement (1), the simulation space is uniformly decomposed into blocks.
In order to support refinement (2), blocks can be further subdivided into equally sized smaller blocks:
$2 \times 2 = 4$ in this 2D illustration and $2 \times 2 \times 2 = 8$ in our 3D simulations.
In (1) as well as in (2), each block is assigned a grid of the same size (last picture of each row).
As a result, transitions between different grid resolutions only occur on the boundary of blocks.
For our massively parallel algorithm for the LBM on nun-uniform grids, exactly this kind of grid allocation strategy is used.
However, the core data structures of \Walberla{} also support arbitrarily sized grids for each block, enabling different grid structures for other kinds of simulations.}
  \label{fig:domain_decomposition}
\end{figure}

In \Walberla{}, the simulation space is decomposed into blocks \cite{Feichtinger2011105}.
These blocks can be assigned arbitrary data, meaning each block can store an arbitrary number of different C++ data structures.
These can be classes provided by the framework, other user-defined classes, STL containers, etc.
In a parallel environment, one block is always assigned to exactly one process.
By assigning a block to a specific process, this process becomes responsible for the part of the simulation space that corresponds to this block.
One block cannot be assigned to multiple processes, but multiple blocks can be assigned to the same process. As a consequence,
these blocks are the smallest entity that can be used for workload distribution, meaning load balancing is achieved by distributing all blocks to all available processes.
For simulations without grid refinement,
the simulation domain is uniformly decomposed into blocks and each block stores a Cartesian grid that corresponds to its part of the simulation domain (see Figure~\ref{fig:domain_decomposition}).
Typically, exactly one block is assigned to each process and ghost layers of the grid on each block are used during the communication in order to synchronize neighboring blocks in parallel simulations (see section~\ref{sec:pconcept:communication}).
Relying on a block-based rather than a cell-based domain partitioning is a parallelization approach that is also adopted by other software frameworks for the LBM~\cite{krause2010,fietz2012}.

In order to achieve our goal of high node level efficiency and scalability to massively parallel systems combined with support for simulations on non-uniform grids,
the domain partitioning concepts described in \cite{Feichtinger2011105} must be extended significantly.
A fully distributed data structure is required for scalability to hundreds of thousands of processes and beyond.
An essential property of such a fully distributed data structure is independence of the per-process memory requirements for managing the data structure itself from the total number of processes.
If two processes are assigned the same number of blocks with the same amount of data that is associated with each block,
both processes must use the same amount of memory during simulation, regardless of whether the simulation runs with hundreds, thousands, or millions of processes.

\begin{figure}[tbp]
  \centering
  \includegraphics[width=0.9\textwidth]{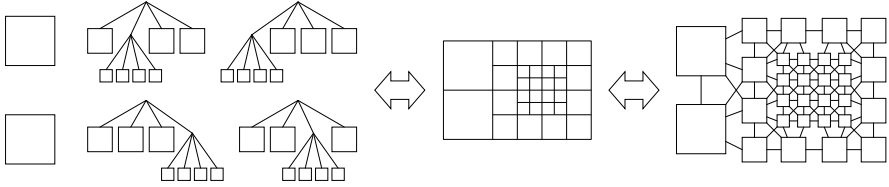}
  \caption{Schematic 2D illustration of the domain partitioning introduced in Figure~\ref{fig:domain_decomposition}(2).
The largest blocks, i.e., the blocks of the initial Cartesian decomposition, act as roots for individual octrees.
As such, they create a forest of octrees visualized on the left (note that quadtrees in this 2D illustration correspond to octrees in 3D).
During simulation, this forest of octrees is not stored explicitly,
but it is defined implicitly by assigning each block a unique block ID that allows to reconstruct the block's exact location within the forest of octrees.
Additionally, every block stores a pair of type (block ID, process rank) for every neighbor,
thereby creating a distributed adjacency graph for all blocks (illustrated on the right).}
  \label{fig:treemesh}
\end{figure}

For simulations that rely on grid refinement, we implemented a distributed forest of octrees data structure that
strictly follows these design principles
and results from further subdividing the blocks of the initial Cartesian domain partitioning into eight equally sized smaller blocks (see Figure~\ref{fig:domain_decomposition}).
Using this data structure, the simulation domain is still partitioned into blocks.
However, blocks now correspond to leaves of the forest of octrees (see Figure~\ref{fig:treemesh}).
As a result, this new domain partitioning structure supports the re-use of existing algorithms.
The developer must usually only define algorithms that work on blocks,
regardless of whether the domain is uniformly partitioned into blocks or a more complex octree partitioning is used.
Only few algorithms that are closely related to how the blocks are arranged in space, as for instance parts of the communication layer (see section~\ref{sec:pconcept:communication}),
require a second implementation tailored to the specifics of the forest of octrees data structure.
If we enforce 2:1 balanced octrees,
the resulting block structure perfectly mirrors the 2:1 balanced grid structure required for the refinement scheme for the LBM described in section~\ref{sec:LBM:refinement}
and as such proves to be a suitable domain partitioning scheme.
At this point we would like to remark that the idea of first partitioning the simulation domain into blocks using an octree approach
and later creating Cartesian meshes inside these blocks
was recently also adopted by a new software project: ForestClaw~\cite{Burstedde2014}.
ForestClaw uses p4est~\cite{Burstedde2011} for domain partitioning, a library that already demonstrated scalability to massively parallel systems
and, being based on a distributed forest of octrees implementation, shares similarities with our approach.

An important, unique feature of our implementation is the fact that every block is aware of all of its spatially adjacent neighbor blocks,
effectively creating a distributed adjacency graph for all blocks (see Figure~\ref{fig:treemesh}).
As a result, we can not only execute algorithms typically associated with octrees but also algorithms that operate on more general graphs.
The spatial partitioning of the simulation domain, however, geometrically always corresponds to a forest of octrees.
Parent nodes or parent-child relationships do not need to be saved explicitly, the tree structure is implicitly defined by block IDs.
Block IDs are used to uniquely identify blocks within the distributed data structure.
They allow to reconstruct a block's exact location within the forest of octrees.
During the simulation, each process only knows about blocks that are stored locally and blocks that are neighboring these process-local blocks.
For all non-local neighbor blocks, this knowledge is only comprised of a globally unique block ID and a corresponding process rank.
Generally, processes have no knowledge about blocks that are not located in their immediate process neighborhood.
As a consequence, the memory usage of a particular process only depends on the number of blocks assigned to this process,
and not on the global number of blocks that are available for the entire simulation.
This kind of fully distributed data structure
enables scalability of simulations working on non-uniform grids to massively parallel systems (see section~\ref{sec:benchmarks}).

\subsection{Two-stage initialization}\label{sec:pconcept:init}

\begin{figure}[tbp]
  \centering
  \resizebox{0.8\textwidth}{!}{\input{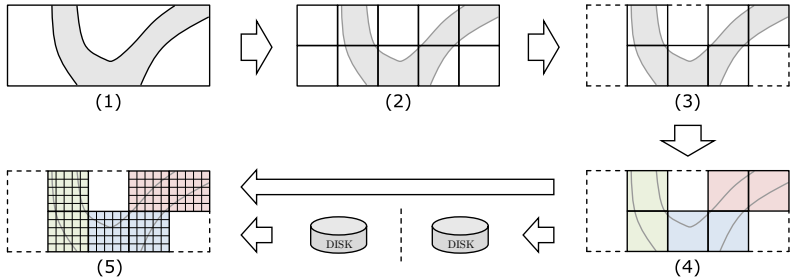}}
  \caption{Initialization starts with a uniform block partitioning (2).
Blocks not needed for the simulation can be discarded (3).
The load balancing step then assigns blocks to processes (4).
If refinement is required, in-between step (3) and (4), some - or all - of the initial blocks are further subdivided into finer blocks (see Figure~\ref{fig:domain_decomposition}).
The resulting domain partitioning can be stored to file in order to be used as the starting point of the actual simulation (5).
If the initialization runs during the actual simulation, step (5) will follow immediately after step (4) without the intermediate storage to file.
In (5), the grid used for the computation is allocated locally on each process.}
  \label{fig:initialization}
\end{figure}

During the initialization, the construction of the data structure starts with a uniform block grid (see Figure~\ref{fig:initialization}).
Each of these initial blocks acts as the root of an octree (see Figure~\ref{fig:treemesh}).
During an iterative process, blocks are then uniformly divided into eight smaller blocks as long as they are marked for refinement by user-defined callback functions.
Our algorithms ensure a 2:1 balance between neighboring blocks by also subdividing blocks
that have not explicitly been marked for refinement by the user or the application.
If no such callback functions are registered, the initial block grid remains unchanged and
can act as a basis for simulations that only need a uniform block partitioning of the simulation space.
Blocks located in parts of the initial space not needed during the simulation can be completely removed from the data structure, 
allowing the block partitioning to handle any kind of geometry and to adapt to arbitrarily shaped domains (see Figure~\ref{fig:initialization}).

During initialization, no actual data is allocated, only two values are stored for each block: memory requirement and workload.
These values are set by a user-defined, application-dependent callback function and are then used in the subsequent load balancing step.
For load balancing, different strategies based on space filling curves and on the graph partitioning library METIS~\cite{karypis1999} are available.
The weight of each block corresponds to the assigned workload.
The task of the load balancing algorithm is to assign a target process ID to each block
such that all blocks are distributed equally among all processes with respect to their weight.
For the balancing strategy that is based on space filling curves, we first
create a linearized array that contains all blocks by traversing all trees in the forest of octrees
and then we divide this array into contiguous pieces of equal weight~\cite{Burstedde2011}.
When using METIS, we transform the adjacency graph containing all blocks (see Figure~\ref{fig:treemesh}) to the graph data structure used by METIS.
Via user-defined callback functions we allow to also specify weights for all block-block connections ($=$ edges in the adjacency graph).
Typically, these weights will correspond to the expected communication volume.
This kind of graph-based load balancing on a block level was also proposed by~\cite{fietz2012}.
For more details on our load balancing process in combination with LBM-based simulations see section~\ref{sec:lbmref:load}.
The end of the load balancing step also marks the end of the initialization phase.
Not needing the entire, fully resolved grid during load balancing is a major advantage and enables very large domains for current, massively parallel supercomputers.
This hierarchical approach of first dividing the simulation domain into blocks and later filling these blocks with corresponding parts of the global grid
allows us to handle computational grids that consist of trillions of cells~\cite{Godenschwager2013}.

The initialization phase can be part of the actual simulation or it can be run prior to the simulation
on a completely different machine with a different number of processes (see Figure~\ref{fig:initialization}).
On a typical desktop computer, the entire initialization phase takes a few milliseconds for several thousands of blocks and,
depending on the chosen load balancing algorithm, several seconds for millions of blocks which are needed for massively parallel simulations.
Running the initialization of the data structure prior to the actual simulation allows
to fine-tune the domain partitioning by trying different load balancing strategies and varying corresponding configuration parameters.
Once this results in a satisfying domain partitioning, the corresponding forest of octrees data structure can be saved to file.
Running the actual simulation then starts by one process reading the file and broadcasting the binary file content to all the other processes.
On a current IBM Blue Gene/Q supercomputer, this process of reading the file, broadcasting its content,
and setting up the forest of octrees data structure on each process according to the information in the file only takes a few seconds
for a simulation involving millions of blocks.
In this particular case, the size of the corresponding file storing the forest of octrees structure is typically in the range of 100 to 200\,MiB.
As a result of this two-stage approach, the actual simulation can start almost immediately and as such contrasts favorably with alternative initialization strategies that, at least initially, require the entire, fully resolved grid for partitioning and therefore can be much more time and memory consuming.

\subsection{Inter-block communication}\label{sec:pconcept:communication}

In a parallel environment, the \Walberla{} framework takes care of the necessary communication between the blocks of the domain partitioning
by providing an extensible, feature rich communication layer that can be adapted to the needs and communication patterns of the underlying simulation.
Generally, the communication is organized into two steps: packing data into and unpacking data from buffers which are exchanged between processes.
If two blocks reside on the same process, the communication layer allows data to be copied directly
without a call to MPI and without the intermediate exchange of a buffer.
For the plain LBM, communication is only performed between spatially adjacent blocks.
For the rest of this article, we refer to two spatially directly adjacent blocks as being connected via a corner if both blocks only intersect in one isolated point.
If both blocks intersect in a line, we refer to them as being connected via an edge.
Analogously, if they intersect in a surface, we refer to them as being connected via a face.

\begin{figure}[tbp]
  \centering
  \includegraphics[width=0.6\textwidth]{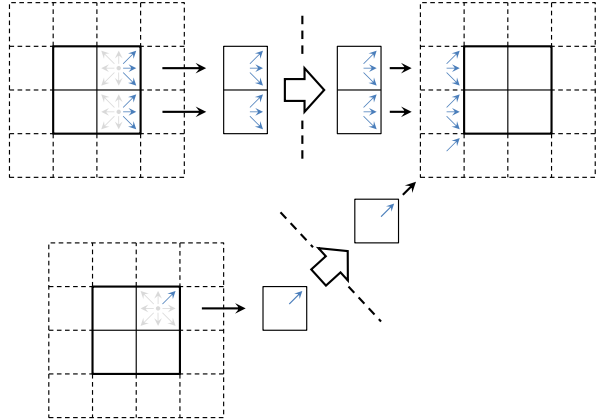}
  \caption{2D example of the LBM with lattice model D2Q9: Every block contains a grid of size $2 \times 2$ with one additional ghost layer.
On edges, three PDF values must be exchanged. On corners, only one PDF value is communicated.}
  \label{fig:communication}
\end{figure}

For parallel simulations with the LBM,
the grid assigned to each block is extended by at least one additional ghost layer that is used to synchronize the data.
During communication, PDF values stored in the outermost inner cells are copied to the corresponding ghost layers of neighboring blocks (see Figure~\ref{fig:communication}).

Depending on the lattice model, exchanging PDF values with fewer neighbors may be sufficient.
If D3Q19 is used, typically, PDF values must be exchanged only with neighboring blocks connected via a face or an edge, but not across a corner.
Additionally, in most simulations, not all PDF values must be communicated, but only those streaming into the neighboring block (see Figure~\ref{fig:communication}).
For D3Q19, out of the 19 PDF values stored in each cell, five PDF values must be exchanged for every cell on a face-to-face connection and only one PDF value for every cell on an edge-to-edge connection.
On average, this reduces the amount of data that must be communicated by a factor of 4.
It should be noted that the modular design of the communication layer allows to adapt all communication steps to the specific needs of the application.
If an application requires more PDF data to be present in ghost layers, always also more data can be exchanged between processes.

\begin{figure}[tbp]
  \centering
  \includegraphics[width=0.7\textwidth]{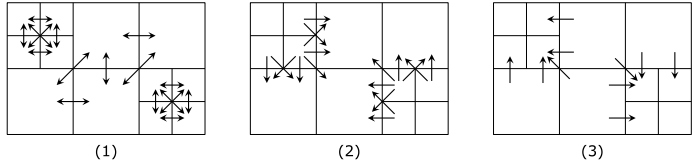}
  \caption{For the forest of octrees data structure, three different communication patterns exist:
data exchange between blocks of the same size (1), fine-to-coarse communication (data packed on small and unpacked on larger block) (2),
and coarse-to-fine communication (data packed on large and unpacked on smaller block) (3).}
  \label{fig:nonuniform_communication}
\end{figure}

For the forest of octrees data structure, we had to implement a new communication scheme that consists of
three different communication patterns (see Figure~\ref{fig:nonuniform_communication}):
(i) data exchange between two blocks of the same size on the same level,
(ii) sending data to a larger block on a coarser level (fine-to-coarse communication),
and (iii) sending data to a smaller block on a finer level (coarse-to-fine communication).
Details on the implementation of these different communication patterns for the LBM on nun-uniform grids follow in sections~\ref{sec:lbmref:equal},~\ref{sec:lbmref:ctf}, and~\ref{sec:lbmref:ftc}.

With and without refinement, always only one message is exchanged between two processes.
This message may contain the data of multiple blocks.
When operating in hybrid OpenMP and MPI mode, these messages can be sent and received in parallel by different threads.
Packing data into buffers prior to sending and unpacking data after receiving can also be done in parallel by different threads if OpenMP is activated.
The OpenMP implementation of the communication module follows a task parallelism model.
First, a unique work package is created for every face, every edge, and every corner of every block.
These work packages are then processed in parallel by all available threads.
As a result, as long as all compute kernels are also parallelized using OpenMP (following a data parallelism model, see section~\ref{sec:lbmref}), all major parts of the simulation are executed thread-parallel.

\section{The LBM on non-uniform grids}\label{sec:lbmref}

In our parallel implementation,
grid refinement for the LBM is based on the forest of octrees block partitioning of the simulation space described in the previous section.
Each block is assigned a grid with the same number of cells in x-, y-, and z-direction (cf.\ Figure~\ref{fig:domain_decomposition}).
As a result, the interfaces between subdomains of different resolution coincide with the block boundaries.

\begin{figure}[tbp]
  \centering
  \includegraphics[width=0.8\textwidth]{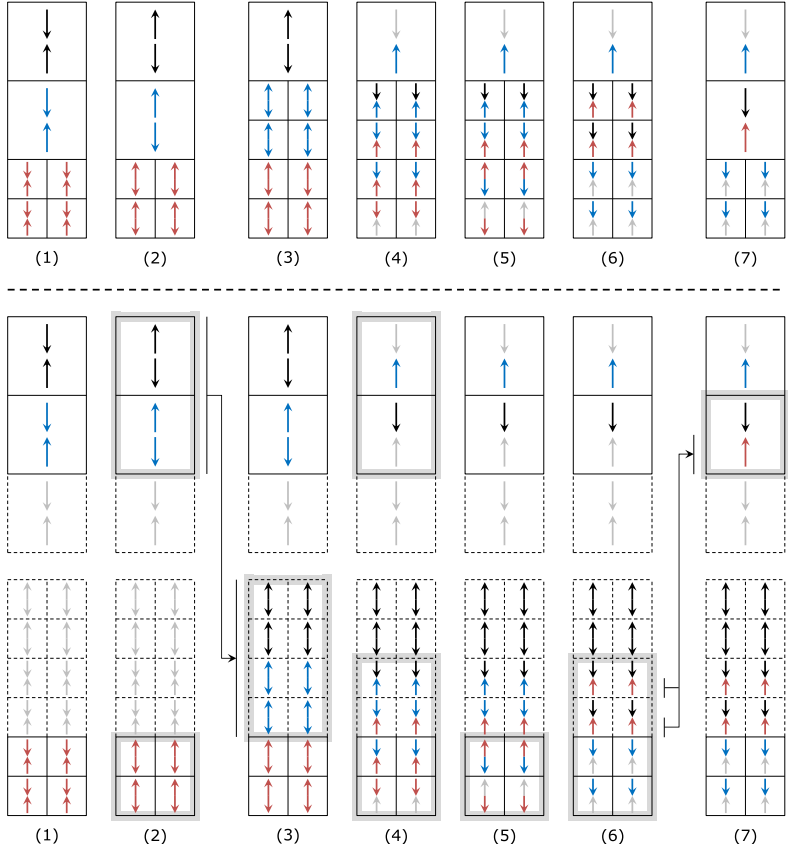}
  \caption{The series of images above the dashed line illustrates one time step on the coarse grid and two corresponding time steps on the fine grid.
The images below the dashed line show the exact same process, but here we illustrate the data distribution of our parallel implementation.
Since grid level transitions coincide with block boundaries, we also visualize cells in ghost layers involved in the algorithm.
Cells that change their content from one step to the next are marked with a gray, surrounding box. Once step (7) is reached, the state of the algorithm is again identical to step (1).}
  \label{fig:algorithm}
\end{figure}

Figure~\ref{fig:algorithm} schematically illustrates the algorithm proposed by \cite{Rohde2006} and our parallel implementation.
Initially (1), the simulation is in a post-streaming state. Our parallel algorithm starts with performing a collision in coarse and fine cells (2).
Cells in ghost layers are not included in the collision.
The content of two coarse cells is then transferred to four ghost layers in the fine grid as indicated by the arrow between step (2) and (3).
Next, streaming is performed in both grids, including the two innermost ghost layers of the fine grid (4).
In step (5) and (6), another collision and streaming step is performed in the fine grid only. Streaming again includes the two innermost ghost layers.
Finally, PDF values in these two innermost ghost layers are merged and transferred to the coarse grid as indicated by the arrow between step (6) and (7).
The implementation and the methods used for transferring data between different grid levels are covered in detail in the following subsections.
Note that from (6) to (7) not all PDF values are transferred from fine to coarse, but only those streaming into the coarse grid (marked red in the illustration).
Also note that the ghost layer of the coarse grid is not required by this algorithm for fine-to-coarse or coarse-to-fine communication.
As a consequence, four ghost layers are only needed for blocks with coarser neighbors. Also, four layers of cells are never communicated.
At most, two layers are transferred during coarse-to-fine communication (see section~\ref{sec:lbmref:ctf}).
For fine-to-coarse communication and communication between two equally sized blocks, significantly less data must be transferred (see sections~\ref{sec:lbmref:ftc} and~\ref{sec:lbmref:equal}).

\subsection{Implementation}\label{sec:lbmref:algo}

\begin{algorithm}[tbp]
\footnotesize
\caption{NonUniformTimeStep}
\label{algo:ref_time_step}
\DontPrintSemicolon
\SetKwFunction{Func}{NonUniformTimeStep}
\SetKwFunction{RefStep}{NonUniformTimeStep}
\SetKwFunction{Collision}{\uwave{CollisionStep}}
\SetKwFunction{Stream}{\uwave{StreamingStep}}
\SetKwFunction{Communication}{\uline{Communication}}
\SetKwFunction{Explosion}{\uline{Explosion}}
\SetKwFunction{Coalescence}{\uline{Coalescence}}
\SetKwProg{Fn}{Function}{}{end}
\Fn{\Func{level $L$}}
{
\lForAll{blocks on level $L$}{\Collision\tcc*[f]{LBM collision step}}
\If{$L \neq \text{finest level}$}
{
recursively call \RefStep{$L+1$}\tcc*[r]{recursive call}
}
\If{$L \neq \text{coarsest level}$}
{
call \Explosion{$L,L-1$}\tcc*[r]{coarse-to-fine communication}
\tcc*[r]{initiated by fine level}
}
call \Communication{$L$}\tcc*[r]{equal-level communication}
\lForAll{blocks on level $L$}{\Stream\tcc*[f]{LBM streaming step}}\label{algo:ref_time_step:1st_stream}
\If{$L \neq \text{finest level}$}
{
call \Coalescence{$L,L+1$}\tcc*[r]{fine-to-coarse communication}
\tcc*[r]{initiated by coarse level}
}
\If{$L = \text{coarsest level}$}
{
\Return\tcc*[r]{end recursion}
}
\lForAll{blocks on level $L$}{\Collision\tcc*[f]{LBM collision step}}\label{algo:ref_time_step:2nd_collide}
\If{$L \neq \text{finest level}$}
{
recursively call \RefStep{$L+1$}\tcc*[r]{recursive call}
}
call \Communication{$L$}\tcc*[r]{equal-level communication}
\lForAll{blocks on level $L$}{\Stream\tcc*[f]{LBM streaming step}}
\If{$L \neq \text{finest level}$}
{
call \Coalescence{$L,L+1$}\tcc*[r]{fine-to-coarse communication}
\tcc*[r]{initiated by coarse level}
}
}
\end{algorithm}

Algorithm~\ref{algo:ref_time_step} displays the basic structure of the recursive procedure that
represents the program flow control for our approach of the LBM on non-uniform grids.
To improve readability, functions that perform the collision or streaming step of the LBM are highlighted with a wavy underline
and functions that involve communication are underlined with a solid line.
The algorithm is called with $L$ equal to zero.
As a result, the simulation is advanced by one coarse time step ($L$ equal to zero corresponds to the coarsest level).
Function $Explosion$ performs coarse-to-fine communication.
In its most basic implementation, PDF values from coarse grid cells are homogeneously distributed to fine cells according to
\begin{equation}
f_{\alpha,fine}(\mathbf{x}_j,t) = f_{\alpha,coarse}(\mathbf{x}_i,t) \label{eq:ctof},
\end{equation}
where $\mathbf{x}_j$ are all the cell centers on the fine grid that correspond to $\mathbf{x}_i$ on the coarse grid.
Function $Coalescence$ performs fine-to-coarse communication.
Here, PDF values stored in the fine grid are transferred to the coarse grid according to
\begin{equation}
f_{\alpha,coarse}(\mathbf{x}_i,t) = \frac{1}{n} \sum\limits_{j=1}^n f_{\alpha,fine}(\mathbf{x}_j,t) \label{eq:ftoc},
\end{equation}
where $n$ is the number of fine cells that correspond to one coarse cell ($n=4$ in 2D and $n=8$ in 3D).
Function $Communication$ takes care of transferring PDF data between grids on the same level.
Boundary conditions are executed as a pre-streaming operation (cf.\ section~\ref{sec:LBM:boundaries}) and are realized at the beginning of function $StreamingStep$.
Both the streaming and collision step are only performed in the interior part of the grid, not in ghost layers, with one exception:
For blocks with at least one larger neighbor block with a coarser grid, the treatment of boundary conditions and streaming also include the two innermost ghost layers.

The actual implementation includes further optimizations as outlined in Algorithm~\ref{algo:ref_time_step_final}.
On the finest grid level, the streaming and collision step (cf.\ lines~\ref{algo:ref_time_step:1st_stream} and~\ref{algo:ref_time_step:2nd_collide} of
Algorithm~\ref{algo:ref_time_step}) can be combined into one fused stream-collide operation (cf.\ line~\ref{algo:ref_time_step_final:stream_collide} of Algorithm~\ref{algo:ref_time_step_final}).
Combining the streaming and collision step significantly reduces the amount of data that is transferred between CPU and main memory.
Not combining both steps requires to load and store the data from and to memory twice.
Fusing both steps allows, for each cell, to read PDF values from neighboring cells (= streaming),
perform the collision step, and store the result back to memory~\cite{CPE:CPE3180}.
Since most time steps are executed on the finest level, the majority of the workload is generated by the blocks on this level.
Thus, improving the performance of the algorithm on the finest level is essential for achieving good overall performance.
Function $StreamCollide$ also includes the treatment of boundary conditions as a pre-streaming step.

\begin{algorithm}[tbp]
\footnotesize
\caption{NonUniformTimeStep (optimized version)}
\label{algo:ref_time_step_final}
\DontPrintSemicolon
\SetKwFunction{Func}{NonUniformTimeStep}
\SetKwFunction{RefStep}{NonUniformTimeStep}
\SetKwFunction{Collision}{\uwave{CollisionStep}}
\SetKwFunction{Stream}{\uwave{StreamingStep}}
\SetKwFunction{StartCommunication}{\uline{InitiateCommunication}}
\SetKwFunction{EndCommunication}{\uline{FinishCommunication}}
\SetKwFunction{StartExplosion}{\uline{InitiateExplosion}}
\SetKwFunction{EndExplosion}{\uline{FinishExplosion}}
\SetKwFunction{StartCoalescence}{\uline{InitiateCoalescence}}
\SetKwFunction{EndCoalescence}{\uline{FinishCoalescence}}
\SetKwFunction{Interpolation}{ExplosionInterpolation}
\SetKwFunction{StreamCollide}{\uwave{StreamCollide}}
\SetKwProg{Fn}{Function}{}{end}
\SetKw{and}{and}
\Fn{\Func{level $L$}}
{
\lForAll{blocks on level $L$}{\Collision}
\If{$L \neq \text{coarsest level}$}
{
call \StartExplosion{$L,L-1$}
}
call \StartCommunication{$L$}\;\label{algo:ref_time_step_final:1st_start_comm}
\If{$L \neq \text{finest level}$}
{
recursively call \RefStep{$L+1$}\tcc*[r]{recursive call}\label{algo:ref_time_step_final:1st_rec_call}
call \StartCoalescence{$L,L+1$}\;
}
\If{$L \neq \text{coarsest level}$}
{
call \EndExplosion{$L,L-1$}\;\label{algo:ref_time_step_final:end_explosion}
call \Interpolation{$L$}\tcc*[r]{interpolation of exploded values}\label{algo:ref_time_step_final:interpolation}
}
call \EndCommunication{$L$}\;\label{algo:ref_time_step_final:1st_end_comm}
\eIf{$L = \text{finest level}$ \and $L \neq \text{coarsest level}$}
{
\tcp{fused LBM streaming \& collision step}
\lForAll{blocks on level $L$}{\StreamCollide}\label{algo:ref_time_step_final:stream_collide}
}
{
\lForAll{blocks on level $L$}{\Stream}
\If{$L \neq \text{finest level}$}
{
call \EndCoalescence{$L,L+1$}
}
\If{$L = \text{coarsest level}$}
{
\Return
}
\lForAll{blocks on level $L$}{\Collision}
}
call \StartCommunication{$L$}\;
\If{$L \neq \text{finest level}$}
{
recursively call \RefStep{$L+1$}\tcc*[r]{recursive call}
call \StartCoalescence{$L,L+1$}\;
}
call \EndCommunication{$L$}\;
\lForAll{blocks on level $L$}{\Stream}
\If{$L \neq \text{finest level}$}
{
call \EndCoalescence{$L,L+1$}
}
}
\end{algorithm}

When hybrid parallel execution with OpenMP is used, functions $CollisionStep$, $StreamingStep$, and $StreamCollide$,
which are always called on an entire block, follow a NUMA-aware data parallelism model,
meaning the work that must be performed for every cell
is distributed evenly among all available threads.
Employing a data parallelism model for the compute kernels for the LBM
when using a block-based domain partitioning is also suggested in~\cite{heuveline2009,krause2010}.

In general, the aforementioned three functions may include any performance optimizations as long as
the employed optimization strategies only affect the current collision, streaming, or fused stream-collide step.
For our simulations, we typically rely on compute kernels for the LBM that make use of a ``structure of arrays'' data layout,
loop splitting to reduce the number of concurrent store streams, and SIMD processing~\cite{CPE:CPE3180}.
Optimization techniques that do not allow separate collision and streaming steps like for example the AA pattern~\cite{5362489},
which requires an adapted collision followed by a fused stream-collide-stream operation,
are not straightforwardly compatible with the presented grid refinement scheme.
As a consequence, we use a two-grid approach for storing the PDF data and performing the streaming step.

As one further step of optimization, communication is executed asynchronously.
In particular, this permits to overlap computation and communication.
To this end, every communication step is separated into two phases.
During the first phase, all message data is packed into a buffer and
the communication is initiated by calling a non-blocking, asynchronous send function of MPI.
Simultaneously, all the matching MPI receive functions are scheduled.
In the second phase, the communication is finalized by waiting for the MPI receive functions to complete.
The actual transfer of data is executed between these two phases.
As a result, when reaching the second phase, the data that has already arrived can be unpacked and computation can continue immediately without stalling.
The first communication between blocks on the same grid level, for example, is initiated in line~\ref{algo:ref_time_step_final:1st_start_comm} of Algorithm~\ref{algo:ref_time_step_final}.
The matching completion of this communication operation happens in line~\ref{algo:ref_time_step_final:1st_end_comm}.
While the corresponding data is transferred between processes, all finer blocks are processed as a result of the recursive call in line~\ref{algo:ref_time_step_final:1st_rec_call},
coalescence is initiated, and explosion is finished.

Finally, the homogeneous distribution of PDF values from one coarse cell to multiple finer cells (cf.~(\ref{eq:ctof})) is improved by a mass and momentum conserving interpolation scheme presented in~\cite{Chen2006}.
After all necessary PDF values were sent from the coarse grid, this interpolation is carried out by function $ExplosionInterpolation$ (cf.\ lines~\ref{algo:ref_time_step_final:end_explosion} and~\ref{algo:ref_time_step_final:interpolation} in Algorithm~\ref{algo:ref_time_step_final}).

The algorithm proposed here allows grid level transitions to be present at fluid-obstacle interfaces if these interfaces are treated with simple bounce back (no slip) or symmetric (free slip) boundary conditions.
To this end, the treatment of boundary conditions is extended to the two innermost ghost layers on fine blocks at the interface to coarse neighbors.
Furthermore, cells marked as fluid or obstacle cells must be marked consistently across the interface region where coarse and fine cells overlap.
If, at the interface between different grid levels, a coarse cell is marked as being a fluid/obstacle cell, all eight corresponding fine cells must also be marked as fluid/obstacle cells.

Even though the asynchronous communication scheme
helps to improve performance, it is still crucial for parallel performance to implement all communication patterns with as little data transfer as possible.
In the following subsections, we will describe how the parallel communication is optimized in the three possible situations, i.e.,
when communication occurs between blocks on the same level, when data is transferred from coarse to fine grids, and vice versa.

\subsection{Communication between blocks on the same level}\label{sec:lbmref:equal}

For the communication between blocks on the same level, we first must check whether or not there exists a larger block with a coarser grid in direct proximity of the two blocks that need to communicate.
We refer to a block $C$ as being in direct proximity of two blocks $A$ and $B$ if there exists a non-empty intersection, $A \cap B \cap C \neq \emptyset$, of all three blocks.
For the remainder of this section, we assume that $A$ and $B$ are adjacent blocks on the same level that need to communicate.
Our distributed data structure stores information about process-local blocks and all blocks in the immediate neighborhood of these local blocks (cf.\ section~\ref{sec:pconcept:octree}).
As a result, checking the neighborhood of a block for the existence of a larger block with a coarser grid is a strictly process-local operation.

If no larger blocks are detected,
i.e., if all blocks $C$ with $A \cap B \cap C \neq \emptyset$ are either smaller blocks with a finer grid or blocks on the same level as $A$ and $B$,
then the same communication scheme as used for simulations without grid refinement is applied.
Both blocks exchange one ghost layer and for each cell only those PDF values that stream into the neighboring block are communicated (cf.\ section~\ref{sec:pconcept:communication} and Figure~\ref{fig:communication}).

\begin{figure}[tbp]
  \centering
  \includegraphics[width=0.55\textwidth]{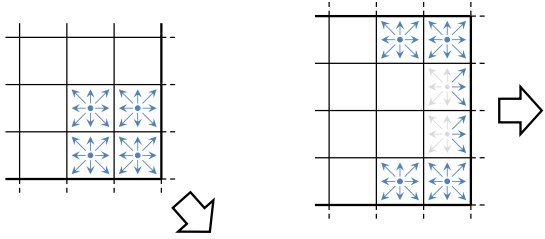}
  \caption{In direct proximity of a larger neighbor block with a coarser grid, for simulations in 2D, blocks on the same level exchange two layers of cells when communicating across a corner (3D:~corners and edges).
When communicating across an edge (3D: face), two layers of cells are only transferred on the corner of the edge.
On the inside of an edge, only one layer is used and only those PDF values that are about to stream into the neighboring block are sent (highlighted in blue).}
  \label{fig:equal_level_communication}
\end{figure}

If, however, there exists a larger block $C$ with $A \cap B \cap C \neq \emptyset$ that contains a coarser grid than blocks $A$ and $B$,
then the two innermost ghost layers of $A$ and $B$ must be included in the streaming step and in the treatment of boundary conditions.
Additionally, we must guarantee that after two consecutive streaming steps on $A$ and $B$, all PDF values in these two ghost layers that are about to be sent to the coarse neighbor are still valid.
Thus, the communication with a neighbor on the same level must be adapted in direct proximity of a larger neighbor block $C$ that contains a coarser grid.
If PDF data is transferred across a corner or an edge, now two layers of cells are communicated, and for each cell, all PDF values are sent.
For communication across a face, however, two layers of cells are only used along the edges of the face. On the inside of the face, only one layer is used.
For these interior cells, only those PDF values that are about to stream into the neighboring block are sent.
This approach is illustrated in Figure~\ref{fig:equal_level_communication} for 2-dimensional simulations (note that faces in 3D correspond to edges in 2D).

In summary, communication between blocks on the same level is, in general, quite similar to the well optimized communication scheme used in simulations without refinement.
Only in direct proximity of larger neighbor blocks with coarser grids, slightly more data must be exchanged.
Since, for 3-dimensional simulations, the amount of data exchanged between blocks is dominated by communication across faces,
ensuring that these transfers operate with the best possible efficiency is crucial for high performance.
In conclusion, communication between blocks on the same level never requires to send more than two layers of cells.
In fact, almost always only one layer of cells that only contains those PDF values streaming into the neighboring block is sent.

\subsection{Coarse-to-fine communication}\label{sec:lbmref:ctf}

\begin{figure}[tbp]
  \centering
  \includegraphics[width=0.7\textwidth]{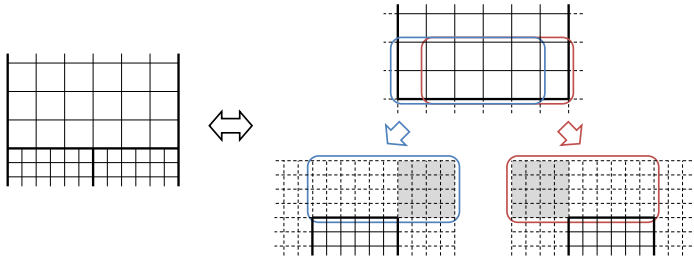}
  \caption{During coarse-to-fine communication, in order to also have valid data in all corners and edges of neighboring fine blocks (cells highlighted in gray),
certain coarse cells must be sent to multiple fine neighbors. Thus, parts of the coarse block that must be transferred will overlap.}
  \label{fig:coarse_to_fine_2}
\end{figure}

\begin{figure}[tbp]
  \centering
  \includegraphics[width=0.7\textwidth]{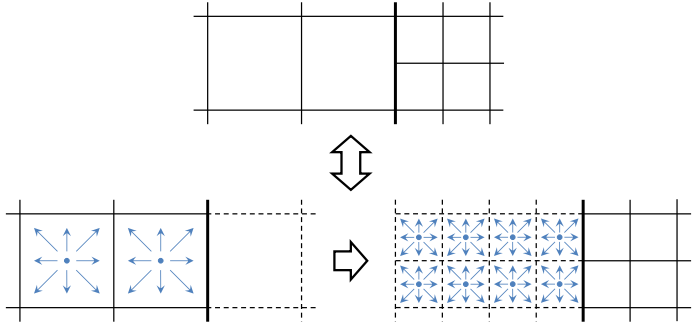}
  \caption{For coarse-to-fine communication, two layers of cells are sent from the interior part of the coarse grid to the ghost layers of the fine grid.
This communication includes all PDF values stored in these cells.}
  \label{fig:coarse_to_fine_1}
\end{figure}

During coarse-to-fine communication, coarse blocks send corner values and the appropriate parts of edges and faces to their finer neighbor blocks.
Some coarse cells must be sent to multiple fine neighbors.
Thus, parts of the coarse block that must be transferred will overlap (see Figure~\ref{fig:coarse_to_fine_2}).
For coarse-to-fine communication, the content of two cell layers is sent from the coarse grid to the fine grid (see Figure~\ref{fig:coarse_to_fine_1}).
On the fine grid, these PDF values are distributed to all four ghost layers.
As mentioned in section~\ref{sec:lbmref:algo}, the implementation of this distribution uses a mass and momentum conserving interpolation scheme from~\cite{Chen2006}.

During coarse-to-fine communication, all PDF values are transferred from the coarse to the fine grid.
Since the treatment of boundary conditions is also performed on the two innermost ghost layers of the fine grid, no PDF values can be omitted.
If only those PDF values are communicated that stream into the fine block, having a simple no slip boundary condition for an obstacle that
spans across a grid level transition would lead to an inconsistent state.
However, if grid level transitions only occur within regions that entirely consist of fluid cells,
sending only one coarse cell instead of two and sending only those PDF values that stream into the fine block is sufficient for accurately performing coarse-to-fine communication.
In careful parallel performance benchmarking,
we did not observe any significant differences in parallel performance since the transfer of data is
mostly hidden by the asynchronous communication and since for most simulations,
the communication time is dominated by communication between blocks on the same level.
This is because the bulk of communication volume is equal-level communication.
Furthermore, since most time steps are performed on the finest grid, the majority of communication takes place between blocks on the finest level.

\subsection{Fine-to-coarse communication}\label{sec:lbmref:ftc}

Fine-to-coarse communication is a post-streaming communication step.
In contrast to coarse-to-fine communication and communication between blocks on the same level,
here, data is read from ghost layers and written into the interior part of the grid (cf.\ steps (6) and (7) in Figure~\ref{fig:algorithm}).
The coarse grid needs information from the two innermost ghost layers of the fine grid.
In order to reduce the amount of data that is communicated,
all the fine cells that correspond to one coarse cell are already aggregated (cf.~(\ref{eq:ftoc})) while packing the data on the sending side.
On the receiving side, i.e., on the coarse grid, these aggregated PDF values are then copied to the appropriate part of the grid.

\begin{figure}[tbp]
  \centering
  \includegraphics[width=0.7\textwidth]{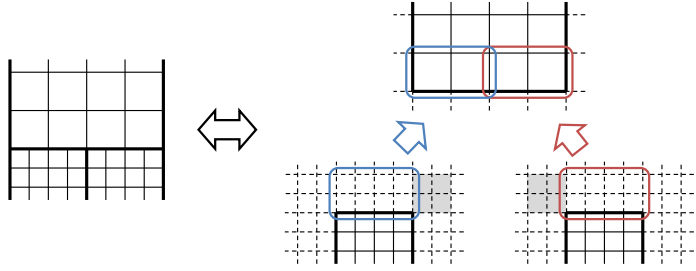}
  \caption{If PDF values are sent from fine to coarse blocks, they must be mapped to the right part of the coarse block.
Cells highlighted in gray never have to be sent.
In contrast to coarse-to-fine communication and communication between blocks on the same level,
here, PDF values are sent from ghost layers of fine blocks to the interior part of coarse blocks.}
  \label{fig:fine_to_coarse_2}
\end{figure}

The data of some fine cells, however, never needs to be transferred to the coarse grid because of another neighboring fine block whose communication already includes this data (see Figure~\ref{fig:fine_to_coarse_2}).
The following rule applies: If a fine block is connected to a coarse block via a face, these two blocks never communicate across edges or corners.
Analogously, if a fine and a coarse block are connected via an edge, there is no communication across corners for these two blocks.
As a result, all connections of image (2) in Figure~\ref{fig:nonuniform_communication} that are \emph{not} available in image (3) do not perform any communication.
In our parallel implementation of the LBM on non-uniform grids, coarse-to-fine as well as fine-to-coarse communication is only performed on connections illustrated in image (3) of Figure~\ref{fig:nonuniform_communication}.

\begin{figure}[tbp]
  \centering
  \includegraphics[width=0.7\textwidth]{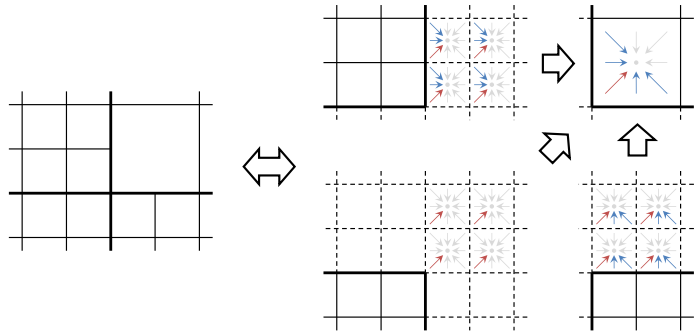}
  \caption{During fine-to-coarse communication, PDF values that originate from the interior part of the coarse block must not be communicated.
Only PDF values that stream into the coarse block (highlighted in red and blue) are sent from ghost layers of fine neighbors.
Some PDF values arrive multiple times from different fine source blocks (highlighted in red).}
  \label{fig:fine_to_coarse_1}
\end{figure}

Also, we must never send all PDF values stored in the fine cells, but only those streaming into the coarse block (see Figure~\ref{fig:fine_to_coarse_1}).
All other PDF values originate from the interior part of the coarse block and must not be overwritten.
Also note that as a result of our parallelization scheme,
some of the PDF values that arrive at the coarse block are sent redundantly from different fine source blocks (highlighted in red in Figure~\ref{fig:fine_to_coarse_1}).

\subsection{Load balancing}\label{sec:lbmref:load}

An even distribution of the workload to all available processes is crucial for good scalability and high performance.
The goal is to prevent processes from being idle while waiting for data from other processes.
For LBM-based simulations, with and without grid refinement,
each block is assigned the same number of lattice cells (cf.\ section~\ref{sec:pconcept:octree} and Figure~\ref{fig:domain_decomposition}).
Distributing these blocks to all available processes such that no process idles during execution is the central task of our load balancing algorithms.

A naive load balancing scheme might look like as follows: Each block is assigned a certain weight.
In LBM-based simulations, this weight must correspond to the block's refinement level.
Since for blocks on a fine level twice as many time steps are executed as for blocks on the next coarser level, 
the weight of a fine block must be twice the weight of an equivalent block on the next coarser level.
As a result, just like the number of time steps, the weight of a block grows exponentially with its level.
Distributing all blocks according to their weights is a perfectly suitable load balancing strategy for LBM-based simulations without grid refinement.

Even if all blocks reside on the same level,
their weights may vary since blocks that only consist of fluid cells typically generate more work than blocks that contain many cells covered by obstacles
which can be skipped by most parts of the algorithm.
Consequently, for plain LBM-based simulations,
setting the block weights to a value that is proportional to the number of fluid cells proves to be a good choice.
Since the weight of each block is calculated by a user-defined callback function,
also more complex schemes that also take into account the work generated by non-fluid cells as proposed by~\cite{fietz2012}
for a simulation of parts of the human respiratory system can be employed.
Using the approach of relating the weight of a block only to the number of fluid cells,
high performance and excellent scalability for a simulation of the human coronary artery tree
on current supercomputers has been demonstrated in~\cite{Godenschwager2013}.
This approach works well for LBM-based simulations without grid refinement since for each time step, all blocks can be processed completely independently.
Synchronization is only necessary either at the beginning or at the end of each time step.

However, perfectly distributing blocks according to their weight is not feasible anymore for a parallel LBM on non-uniform grids.
Here, blocks that reside on different levels cannot be processed completely independently.
At fixed stages during the algorithm, blocks of different levels must interact with each other via fine-to-coarse and coarse-to-fine communication.
For best performance, a suitable load balancing scheme must take into account the structure of the algorithm including all points of communication.

The load balancing scheme that best fits our algorithm processes all levels separately.
For each level, all blocks that reside on this level are distributed to all available processes.
Typically, most of the work is generated by the finest grid levels.
As a consequence, perfectly distributing the work generated by these levels is crucial for maximal performance.
In principle, many load balancing algorithms are suitable for this kind of problem and even a different load balancing algorithm might be used for each level.
In practice, we do not vary the algorithm from level to level.
Currently, we use algorithms that
are either based on space-filling curves using Morton or Hilbert order or that make use of the graph partitioning library METIS \cite{karypis1999} (see section~\ref{sec:pconcept:init}).
Even though blocks on different levels are distributed separately, blocks on the same level can still have different weights due to different fluid to obstacle cell ratios.
The resulting different weights are taken into account during distribution of these blocks to all available processes.
As a result, process idle times are minimized since our level-wise load balancing scheme perfectly fits the structure of the parallel algorithm for the LBM on non-uniform grids.
The need for level-wise load balancing was also pointed out in \cite{Hasert:229088}, but has not yet been implemented.

\section{Validation and benchmarks}\label{sec:benchmarks}

In the following subsections, we present two validation scenarios that verify the accuracy and correctness of our parallelization approach for the algorithm proposed by~\cite{Rohde2006}.
Additionally, we evaluate the performance of our implementation in order to
confirm that our data structures and algorithms are fully capable of efficiently performing massively parallel simulations.
For an extensive discussion on the general accuracy of the method, we refer to the validation in~\cite{Rohde2006}.
We compute two different Poiseuille flow scenarios
in order to verify the correctness of our parallel implementation.
In contrast to~\cite{Rohde2006}, we use the TRT model instead of the SRT model in all our simulations.
Concerning the two relaxation parameters of the TRT model, we propose to keep $\Lambda_{eo}$ (see~(\ref{eq:LBM:lambda_eo})) constant on all levels,
scale $\lambda_e$ according to $\omega$ in~(\ref{eq:LBM:omega_scaling_0}) and~(\ref{eq:LBM:omega_scaling}),
and calculate $\lambda_o$ from $\Lambda_{eo}$ and $\lambda_e$ according to~(\ref{eq:LBM:lambda_oL}) (see section~\ref{sec:LBM:refinement}).

\begin{figure}[tbp]
  \centering
  \includegraphics{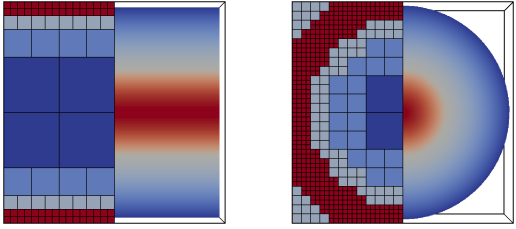}
  \caption{Domain partitioning and flow profiles in the zy-plane for simulations of plane (left) and pipe (right) Poiseuille flow
with grid refinement and four different grid levels.
Only the block partitioning is shown. During simulation, each block consists of $10 \times 10 \times 10$ cells.
In the flow profile, high velocities are depicted in red, zero velocities in dark blue.}
  \label{fig:benchmark:setups}
\end{figure}

In order to enable comparability of different simulations,
velocities are scaled during evaluation such that the maximum velocity given by the corresponding analytical solution is always equal to 1.
For all simulations, $\Lambda_{eo}$ is set to \sfrac{3}{16}, the Reynolds number is fixed to 10,
and the simulation space spans $[0,1] \times [0,1] \times [0,1]$, which is equal to the fluid domain
except for the pipe Poiseuille flow scenario where only a cylindrical part of the simulation space is covered by the fluid domain.
As a consequence, the surface area of any cut perpendicular to the x-axis is either 1 for simulations with a rectangular or $\sfrac{1}{4} \cdot \pi$ for simulations with a circular cross section (cf.\ Figure~\ref{fig:benchmark:setups}).
The volumetric flow rate $Q$ is calculated as
\begin{equation*}
Q = \bar{u} \cdot A,
\end{equation*}
with $\bar{u}$ denoting the mean flow velocity through the cross-sectional area $A$.
In our volumetric approach for the LBM, flow velocities are evaluated at cell centers.
We calculate weighted discrete $L^1$, $L^2$, and $L^\infty$ norms of the error in the velocity by iterating all fluid cells in the domain $\Omega$ and
\begin{equation*}
L^1 = \sum\limits_{i} w_i \cdot \Delta u_i~~,~~L^2 = \sqrt{ \sum\limits_{i} w_i \cdot (\Delta u_i)^2 }~~,~~L^\infty = \max_i \Delta u_i~,
\end{equation*}
where $w_i = (\Delta x_i)^3$ is a grid level-dependent weighting factor that is equal to the volume of cell $i$ and $\Delta u_i = \|\mathbf{u}_i - \mathbf{u}_{i,exact}\|$,
with $\mathbf{u}_i$ denoting the flow velocity computed by the simulation and
$\mathbf{u}_{i,exact}$ denoting the corresponding velocity given by the analytical solution.

Since entire cells are marked as either fluid or obstacle and we employ bounce back boundary conditions, boundaries are always located halfway in-between two cells.
All simulations are run in 3D and make use of the D3Q19 lattice model.

\subsection{Poiseuille flow}\label{sec:benchmarks:poiseuille}

For the simulation of Poiseuille flow,
we drive the fluid with a body force caused by a constant acceleration (cf.\ section~\ref{sec:LBM:forces}) that induces a flow in x-direction.
The acceleration is scaled to different grid levels using~(\ref{eq:LBM:force_scaling}).
All walls are fixed and set to no slip boundary conditions.
We expect a parabolic flow profile that is in accordance with Poiseuille's law.

We use two different setups (cf.\ illustration in Figure~\ref{fig:benchmark:setups}).
By having no slip bounce back boundary conditions at $y=0$ and $y=1$ and periodic boundaries in x- as well as z-direction, we simulate plane Poiseuille flow.
For pipe Poiseuille flow, we use a circular profile in the yz-plane and choose periodic boundaries in x-direction.

\begin{figure}[tbp]
   \centering
   \scalebox{.78}{\input{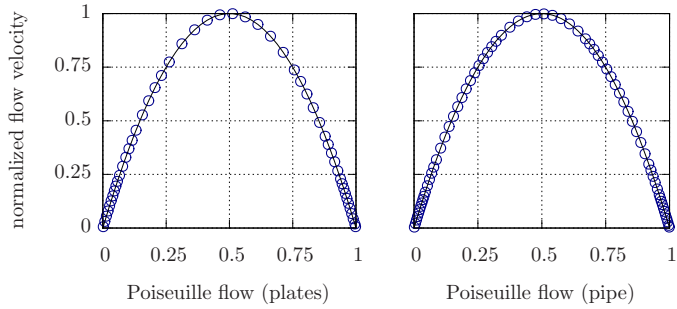}}
   \caption{Velocity profiles evaluated along a line through the middle of the channel, parallel to the y-axis.
The black line follows the analytical solution, the blue circles correspond to the velocities calculated by the simulation (only every 2nd data point is plotted).
The graphs correspond to the two simulations depicted in Figure~\ref{fig:benchmark:setups}.}
  \label{fig:benchmark:accuracy}
\end{figure}

\begin{figure}[tbp]
   \centering
   \scalebox{.78}{\input{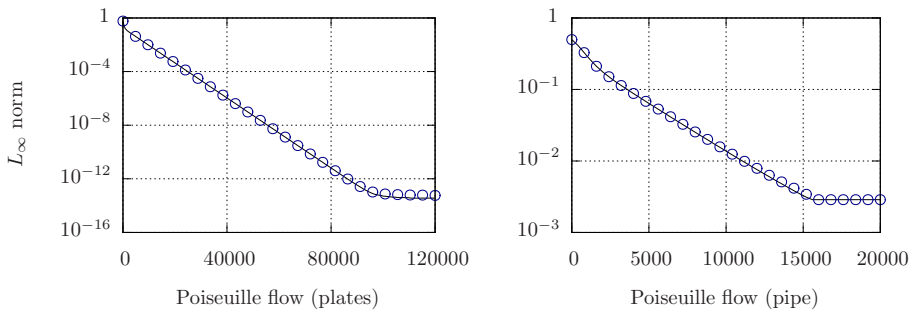}}
   \caption{$L^\infty$ evaluated over the time of the simulation (simulation time in time steps on the finest grid).
The black line corresponds to a simulation with the entire domain refined to the finest resolution.
The blue circles correspond to a simulation with grid refinement with four different grid resolutions and the finest level only at fluid-boundary interfaces
(cf.\ illustration of Figure~\ref{fig:benchmark:setups}).
We see the same agreement when evaluating $L^1$ and $L^2$ norms.}
  \label{fig:benchmark:convergence}
\end{figure}

For plane Poiseuille flow, the flow rate $Q = \bar{u} \cdot A = \sfrac{2}{3} \cdot u_{max} \cdot A = \sfrac{2}{3} \cdot u_{max}$.
Using the TRT model and $\Lambda_{eo}$ equal to \sfrac{3}{16}, we expect the simulation to be accurate on any uniform grid.
Without grid refinement, the simulation converges to the analytical solution, i.e.,
the volumetric flow rate as well as the flow velocities evaluated at all cell centers match the analytical solution once the steady state is reached.
If we add grid refinement, we can retain this convergence behavior to the steady state solution over time.
When refining only at the plate on the bottom ($y=0$), or at the plate on the top ($y=1$), or in the middle between both plates, or at the bottom as well as at the top plate,
convergence over time remains unchanged and we see convergence to the analytical solution, i.e., $L^\infty$ will drop to values close to machine accuracy (cf.\ first graph in Figures~\ref{fig:benchmark:accuracy} and~\ref{fig:benchmark:convergence}).

\begin{figure}[tbp]
   \centering
   \scalebox{.78}{\input{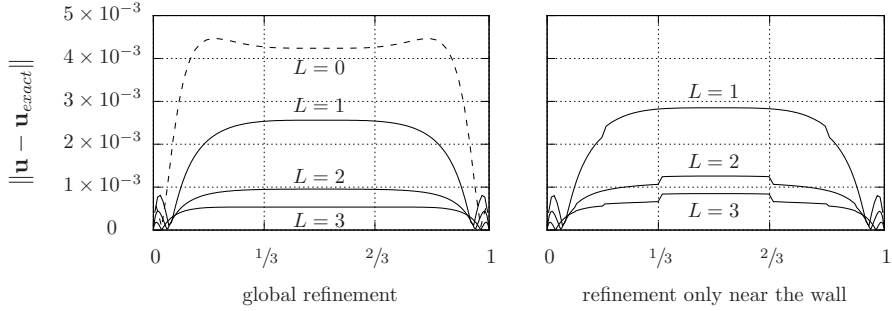}}
   \caption{The error $\|\mathbf{u} - \mathbf{u}_{exact}\|$ of simulations of Poiseuille flow through a channel with circular cross section
evaluated along a line through the middle of the channel, parallel to the y-axis.
The first graph shows the error for simulations where the entire domain is refined either to level 0, 1, 2, or 3.
The second graph shows the error for simulations where the finest level (1, 2, or 3) is only used to resolve the fluid-wall interface.
Towards the center of the channel, the grid resolution becomes increasingly coarser, always reaching level 0 at the center
(cf.\ third illustration in Figure~\ref{fig:benchmark:setups}).
Small jumps in the velocity that are noticeable in the graph on the right are due to errors at the refinement interface and
conform with the observations reported in~\cite{Rohde2006}.}
   \label{fig:poiseuillen_graph}
\end{figure}

For pipe Poiseuille flow with a circular cross section,
the volumetric flow rate $Q = \bar{u} \cdot A = \sfrac{1}{2} \cdot u_{max} \cdot A = \sfrac{1}{8} \cdot u_{max} \cdot \pi$.
For the LBM with no slip conditions at curved boundaries, the expected convergence order is discussed in, e.g.,~\cite{Ginzburg2008}.
Similarly, the accuracy at the refinement interface is studied in~\cite{Rohde2006}.
For this article, the circular profile is approximated by a staircase
and we use the well-established mass-conserving first order bounce back condition,
and thus expect errors that converge linearly with the resolution of the simulation.
When uniformly refining the global grid,
we observe the expected behavior.
Errors decrease linearly the higher the resolution of the underlying grid,
i.e., the more cells we use for modeling the flow inside the pipe (cf.\ ``global refinement'' in Table~\ref{fig:poiseuillen_table}).
Grid refinement can help to avoid resolving the entire domain with a finer grid.
If we use finer grids only near the wall of the pipe (see second illustration in Figure~\ref{fig:benchmark:setups}),
convergence over time remains unchanged (cf.\ second graph in Figure~\ref{fig:benchmark:convergence}) and
we obtain comparable accuracy with simulations that uniformly resolve the entire pipe with fine grids
(see Figure~\ref{fig:poiseuillen_graph}).
Refining only the center of the pipe has no impact on accuracy.
In conclusion, the accuracy of the simulation
is mainly determined by the resolution of the grid used to resolve the fluid-wall interface at the inside wall of the pipe.

Results for all of these simulations of pipe Poiseuille flow are summarized in Table~\ref{fig:poiseuillen_table}.
Accuracy is evaluated after convergence of the flow to a steady state.
For calculating the flow rate $Q$, we evaluate the flow through a cross sectional area in the middle of the channel parallel to the yz-plane.
The relative error compared to the analytical solution of $Q$ is shown in the second column of Table~\ref{fig:poiseuillen_table}.
$L^1$, $L^2$, and $L^\infty$ norms are listed in columns 3, 4, and 5.

\begin{table}[tbp]
  \centering
  \footnotesize
  \caption{Comparison of the accuracy of different simulations of pipe Poiseuille flow.
The simulation was executed with different resolutions and with refining the entire domain to a certain level (global refinement),
with grid refinement to different levels only close to the wall of the pipe (wall refinement),
and with grid refinement to different levels only in the center of the pipe (center refinement).
The corresponding refinement level is stated in brackets.
On the coarsest level, the pipe consists of 60 cells in diameter - leading to 120, 240, and 480 cells for level 1, 2, and 3, respectively.
The linear decrease of the error conforms with the first order bounce back boundary conditions in use.}
  \label{fig:poiseuillen_table}
  \begin{tabular}{rcccc}
  \toprule
  & flow rate error & $L^1$ & $L^2$ & $L^\infty$ \\
  \midrule
  global refinement (0) & $7.96 \times 10^{-3}$ & $3.48 \times 10^{-3}$ & $4.28 \times 10^{-3}$ & $19.2 \times 10^{-3}$ \\
  global refinement (1) & $4.98 \times 10^{-3}$ & $2.06 \times 10^{-3}$ & $2.43 \times 10^{-3}$ & $8.92 \times 10^{-3}$ \\
  global refinement (2) & $1.86 \times 10^{-3}$ & $0.77 \times 10^{-3}$ & $0.90 \times 10^{-3}$ & $4.59 \times 10^{-3}$ \\
  global refinement (3) & $1.05 \times 10^{-3}$ & $0.42 \times 10^{-3}$ & $0.50 \times 10^{-3}$ & $2.87 \times 10^{-3}$ \\
  \midrule
  wall refinement (1) & $5.14 \times 10^{-3}$ & $2.17 \times 10^{-3}$ & $2.54 \times 10^{-3}$ & $8.92 \times 10^{-3}$ \\
  wall refinement (2) & $1.96 \times 10^{-3}$ & $0.84 \times 10^{-3}$ & $0.98 \times 10^{-3}$ & $4.59 \times 10^{-3}$ \\
  wall refinement (3) & $1.13 \times 10^{-3}$ & $0.49 \times 10^{-3}$ & $0.57 \times 10^{-3}$ & $2.87 \times 10^{-3}$ \\
  \midrule
  center refinement (1) & $7.95 \times 10^{-3}$ & $3.46 \times 10^{-3}$ & $4.27 \times 10^{-3}$ & $19.2 \times 10^{-3}$ \\
  center refinement (2) & $7.92 \times 10^{-3}$ & $3.40 \times 10^{-3}$ & $4.21 \times 10^{-3}$ & $19.2 \times 10^{-3}$ \\
  center refinement (3) & $7.92 \times 10^{-3}$ & $3.39 \times 10^{-3}$ & $4.20 \times 10^{-3}$ & $19.2 \times 10^{-3}$ \\
  \bottomrule
  \end{tabular}
\end{table}

\subsection{Performance evaluation}\label{sec:benchmarks:hpc}

In order to evaluate the performance of massively parallel simulations, we run benchmarks on two petascale supercomputers:
JUQUEEN, an IBM Blue Gene/Q system ranked 11th in the TOP500 list\footnote{November 2015},
and SuperMUC, a x86-based system build on Intel Xeon CPUs ranked 23rd in the same list.
JUQUEEN provides 458,752 PowerPC A2 processor cores running at 1.6\,GHz, with each core capable of 4-way multithreading.
Based on our observations in~\cite{Godenschwager2013}, in order to achieve maximal performance,
we make full use of multithreading on JUQUEEN by placing either four processes or four threads of the same process on one core.
The SuperMUC system features fewer, but more powerful, cores than JUQUEEN.
It is built out of 18,432 Intel Xeon E5-2680 processors running at 2.7\,GHz, which sums up to a total of 147,456 cores.
Due to the installation of a hardware upgrade\footnote{January 2015 until summer 2015},
the simulations for this article are limited to 4,096 CPUs (32,768 cores).
On both systems, we use double-precision floating-point arithmetic and the highest compiler optimization level available:
level 5 for the IBM XL compiler on JUQUEEN and level 3 for the Intel compiler on SuperMUC.

For evaluating weak and strong scaling performance of our parallel algorithm for the LBM on non-uniform grids,
we make use of a synthetic benchmark that simulates lid-driven cavity flow in 3D with $\Omega = [0,1] \times [0,1] \times [0,1]$.
The benchmark uses a velocity bounce back boundary condition at $z=1$ in order to simulate the moving lid.
For all other domain boundaries, we use no slip bounce back boundary conditions.
The regions where the moving lid meets the stationary domain boundaries are refined three times.
The properties of the resulting refinement structure are listed in Table~\ref{fig:lid_driven_setup}; they remain fixed for all performance benchmarks.

For the evaluation of weak scaling performance, we assign four blocks of the finest level to every process.
As a result, each process additionally holds one or two blocks of level 2, one or no block of level 1, and one or no block of the coarsest level.
On average, each process will hold 6.6875 blocks, independent of the total number of processes.
When increasing the number of processes, the number of cells per block (and process) remains constant.
As a consequence, the global number of cells increases linearly with the number of processes.
For the evaluation of strong scaling performance, only one block of the finest level is assigned to each process.
As a result, each process additionally holds either one or no block from the other levels.
On average, each process will hold 1.6719 blocks.
The more processes are used, the fewer cells are allocated per block (and process).
As a consequence, the global number of cells remains constant, independent of the total number of processes.

\begin{table}[tbp]
  \centering
  \footnotesize
  \caption{The memory requirements and workload of each grid level as well as the amount of space covered by every level in the lid-driven cavity performance benchmark.
The memory requirements are proportional to the number of blocks (the finest level accounts for 59.81\,\% of all blocks).}
  \label{fig:lid_driven_setup}
  \begin{tabular}{rcccc}
  \toprule
  & $L=0$ & $L=1$ & $L=2$ & $L=3$ \\
  \midrule
  domain coverage ratio & \tablenum[table-format=2.2]{77.78}\,\% & \tablenum[table-format=2.2]{16.67}\,\% & \tablenum[table-format=2.2]{ 4.17}\,\% & \tablenum[table-format=2.2]{ 1.39}\,\% \\
  workload share        & \tablenum[table-format=2.2]{ 1.10}\,\% & \tablenum[table-format=2.2]{ 3.76}\,\% & \tablenum[table-format=2.2]{15.02}\,\% & \tablenum[table-format=2.2]{80.13}\,\% \\
  memory/block share    & \tablenum[table-format=2.2]{ 6.54}\,\% & \tablenum[table-format=2.2]{11.22}\,\% & \tablenum[table-format=2.2]{22.43}\,\% & \tablenum[table-format=2.2]{59.81}\,\% \\
  \bottomrule
  \end{tabular}
\end{table}

We report number of processor cores, not the number of processes.
For a fixed number of cores, we can either run the benchmark with $\alpha$ processes (MPI only) or
run a hybrid simulation with $\frac{\alpha}{\beta}$ processes and $\beta$ OpenMP threads per process.
On SuperMUC, we choose $\alpha$ to be equal to the number of cores.
Since we make full use of multithreading on JUQUEEN, we choose $\alpha$ to be equal to four times the number of cores when running the benchmark on JUQUEEN.
We compare the performance of simulations that rely on MPI only to the performance of hybrid simulations that use four ($\beta=4$) or eight ($\beta=8$) threads, respectively.
Since hybrid simulations use $\beta$ times fewer processes, we allocate $\beta$ times more cells for each block (and process).
As a result, for a given benchmark scenario and for a fixed number of processor cores (= fixed amount of hardware resources),
the amount of work remains constant, independent of the parallelization strategy in use.

For the weak scaling benchmark, we report \MLUPSPS{}, which stands for ``million lattice cell updates'' per second.
For ideal weak scaling, \MLUPSPS{} must scale linearly with the number of processor cores.
For the strong scaling benchmark, we report the number of time steps executed each second on the finest grid.
The more time steps are executed each second, the higher the throughput of the simulation and the shorter the time to solution for a fixed problem size.
For both benchmarks, we also report \MLUPSPS{} divided by the number of processor cores.
When increasing the number of cores, \MLUPSPSC{} is proportional to parallel efficiency.
If \MLUPSPSC{} remains constant, parallel efficiency is equal to 1.
For all measurements, we use the same compute kernels as described in~\cite{Godenschwager2013}.
These compute kernels make use of SIMD instructions on the IBM Blue Gene/Q architecture of JUQUEEN as well as the x86-based hardware of SuperMUC.
Since, in \Walberla{}, the performance of the fastest TRT kernel is identical to the performance of the fastest SRT kernel~\cite{Godenschwager2013},
all results reported here apply for both collision schemes, TRT and SRT.

\begin{figure}[tbp]
   \centering
   \scalebox{.79}{\input{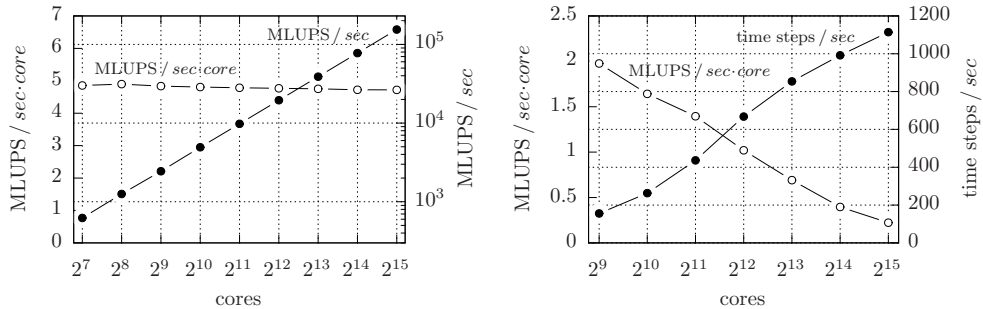}}
   \caption{Weak (left) and strong (right) scaling performance on SuperMUC.
For evaluating weak scaling performance, the benchmark uses 3.34 million cells/core. For evaluating strong scaling performance, the global number of cells is fixed to 8.56 million.
Only the best results of three different parallelization strategies (MPI only, MPI+OpenMP with 4 threads/process, MPI+OpenMP with 8 threads/process) are plotted.
Detailed results are listed in Table~\ref{fig:benchmark:table_supermuc}.}
   \label{fig:benchmark:supermuc}
\end{figure}

For evaluating weak scaling performance on SuperMUC, we use 3.34 million cells per core.
As a consequence, with 32,768 cores of SuperMUC, we can execute simulations that consist of up to 110 billion cells.
For these simulations, we achieve up to 154,599 \MLUPSPS{}.
Since the algorithms do not involve global communication and since our underlying data structures are fully distributed to all processes,
we observe almost perfect weak scaling (see first graph in Figure~\ref{fig:benchmark:supermuc}).
Hybrid simulations using four threads per process are slightly faster than simulations that rely on MPI only and simulations that use eight threads per process (see Table~\ref{fig:benchmark:table_supermuc}).
For the evaluation of strong scaling performance, we fix the global number of cells to 8.56 million and run the same simulation on 512 up to 32,768 cores.
In the largest simulation with 32,768 cores, each core is responsible for only 261 cells.
As expected, parallel efficiency decreases when more processes are used and when fewer cells are assigned to each core.
However, throughput constantly increases up to 1,114 time steps per second on the finest grid (see second graph in Figure~\ref{fig:benchmark:supermuc}),
enabling simulations that perform up to 4 million time steps per hour compute time.
Hybrid simulations are generally faster (cf.~Table~\ref{fig:benchmark:table_supermuc}) since for hybrid simulations fewer processes with larger blocks per process can be used.
As a result, the amount of data communicated between neighboring processes is smaller compared to simulations that use MPI parallelization only.

\begin{table}[tbp]
  \centering
  \footnotesize
  \caption{Detailed results for weak scaling (3.34 million cells/core) and strong scaling (8.56 million cells in total) performance on SuperMUC.
The table lists results for simulations that rely on MPI only and for hybrid simulations with four and eight threads per process.}
  \label{fig:benchmark:table_supermuc}
  \begin{tabular}{rcccccc}
  \toprule
  & \multicolumn{3}{c}{weak scaling (MLUPS\,/\,$sec\!\cdot\!core$)} & \multicolumn{3}{c}{strong scaling (time steps\,/\,$sec$)} \\
  \cmidrule(rl){2-4}
  \cmidrule(rl){5-7}
  cores & MPI only & OpenMP (4) & OpenMP (8) & MPI only & OpenMP (4) & OpenMP (8) \\
  \midrule
  128   & 4.80022 & 4.85441 & 4.75645 & & & \\
  256   & 4.77671 & 4.89659 & 4.64531 & & & \\
  512   & 4.76295 & 4.83518 & 4.63247 & \tablenum[table-format=4.2]{ 127.04} & \tablenum[table-format=4.2]{ 155.34} & \tablenum[table-format=4.2]{ 145.64} \\
  1024  & 4.75190 & 4.80719 & 4.64152 & \tablenum[table-format=4.2]{ 191.40} & \tablenum[table-format=4.2]{ 263.06} & \tablenum[table-format=4.2]{ 248.78} \\
  2048  & 4.72467 & 4.78511 & 4.63214 & \tablenum[table-format=4.2]{ 289.64} & \tablenum[table-format=4.2]{ 422.42} & \tablenum[table-format=4.2]{ 436.48} \\
  4096  & 4.68737 & 4.76766 & 4.63984 & \tablenum[table-format=4.2]{ 412.57} & \tablenum[table-format=4.2]{ 650.54} & \tablenum[table-format=4.2]{ 667.65} \\
  8192  & 4.71740 & 4.75104 & 4.60749 & \tablenum[table-format=4.2]{ 593.36} & \tablenum[table-format=4.2]{ 853.99} & \tablenum[table-format=4.2]{ 775.05} \\
  16384 & 4.64417 & 4.72311 & 4.60263 & \tablenum[table-format=4.2]{ 799.10} & \tablenum[table-format=4.2]{ 991.72} & \tablenum[table-format=4.2]{ 822.86} \\
  32768 & 4.62946 & 4.71798 & 4.57990 & \tablenum[table-format=4.2]{1057.57} & \tablenum[table-format=4.2]{1113.97} & \tablenum[table-format=4.2]{1018.27} \\
  \bottomrule
  \end{tabular}
\end{table}

\begin{figure}[tbp]
   \centering
   \scalebox{.79}{\input{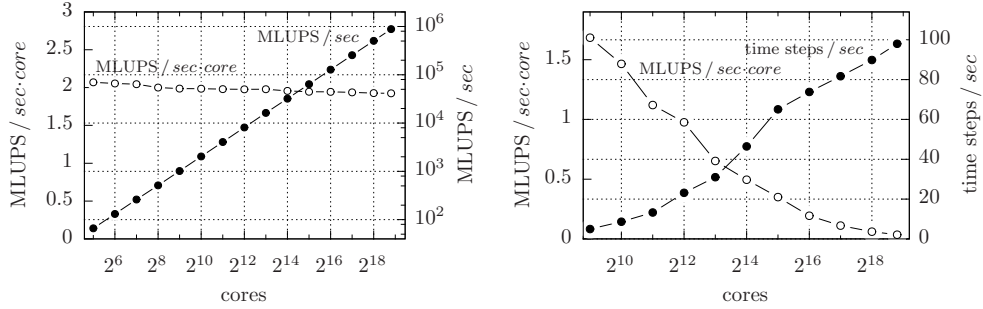}}
   \caption{Weak (left) and strong (right) scaling performance on JUQUEEN.
For evaluating weak scaling performance, the benchmark uses 1.93 million cells/core. For evaluating strong scaling performance, the global number of cells is fixed to 233 million.
Only the best results of three different parallelization strategies (MPI only, MPI+OpenMP with 4 threads/process, MPI+OpenMP with 8 threads/process) are plotted.
Detailed results are listed in Table~\ref{fig:benchmark:table_juqueen}.}
   \label{fig:benchmark:juqueen}
\end{figure}

On JUQUEEN, we observe qualitatively the same behavior.
However, in contrast to SuperMUC, JUQUEEN provides many more processor cores.
The simulations that run on the entire machine make use of 1.835 million threads.
Since these large simulations require data structures that consist of millions of blocks,
we make use of our two-stage initialization process and load
the results of the initialization phase from file instead of creating the domain partitioning and
performing load balancing at runtime (see section~\ref{sec:pconcept:init}).
For the evaluation of weak scaling performance, we use 1.93 million cells per core, resulting in almost one trillion cells (886 billion) for the largest simulation.
Similar to our results on SuperMUC, we observe almost perfect weak scaling (see first graph in Figure~\ref{fig:benchmark:juqueen}),
with hybrid simulations using four threads per process being significantly faster than simulations that rely on MPI only and
simulations that use eight threads per process (see Table~\ref{fig:benchmark:table_juqueen}).
With 889,602 \MLUPSPS{} for the largest simulation on 458,752 cores, we are updating 16.9 trillion PDFs in each second.
For the evaluation of strong scaling performance, we fix the global number of cells to 233 million and run the same simulation on 512 up to 458,752 cores.
In simulations on 458,752 cores, each core is responsible for only 508 cells.
Again, we observe parallel efficiency decreasing when more processes are used and when fewer cells are assigned to each core.
Throughput, however, constantly increases up to 98 time steps per second on the finest grid (see second graph in Figure~\ref{fig:benchmark:juqueen}).
In order to achieve maximal throughput, we must use a hybrid parallelization approach (cf.~Table~\ref{fig:benchmark:table_juqueen}).

Compared to benchmarks for simulations on uniform grids~\cite{Godenschwager2013},
our non-uniform LBM scheme that uses statically refined grid structures looses a factor of 2 to 2.5 in terms of \MLUPSPS{} numbers.
This is due to the fact that in the uniform LBM scheme, there is significantly less overhead.
The uniform scheme does not need to perform interpolation, it communicates less data,
and it can always make use of fused stream-collide compute kernels.
In~\cite{Godenschwager2013}, strong scaling performance is only reported for a simulation of blood flow in the coronary artery tree.
In this kind of simulation,
blocks are less densely populated with fluid cells, resulting in less data that needs to be processed and communicated.
Throughput in this type of simulation is typically two to three times higher than in simulations with blocks that are densely populated with fluid cells.
When comparing the number of time steps that can be executed in one second by our non-uniform LBM scheme to the throughput in uniform simulations,
we observe a decrease by a factor of 3 to 4 for simulations with blocks that are densely populated with fluid cells.
However, this loss in efficiency is more than compensated by the fact that,
in general, non-uniform simulations based on locally refined grid structures generate significantly
less workload and require much less memory than simulations that uniformly resolve the entire domain with the highest resolution.
With more than 1,000 time steps per second,
our benchmark with 4 different levels of refinement achieves considerably higher throughput rates on SuperMUC than on JUQUEEN.
This is also in good agreement with previous observations in~\cite{Godenschwager2013}.

\begin{table}[tbp]
  \centering
  \footnotesize
  \caption{Detailed results for weak scaling (1.93 million cells/core) and strong scaling (233 million cells in total) performance on JUQUEEN.
The table lists results for simulations that rely on MPI only and for hybrid simulations with four and eight threads per process.}
  \label{fig:benchmark:table_juqueen}
  \begin{tabular}{rcccccc}
  \toprule
  & \multicolumn{3}{c}{weak scaling (MLUPS\,/\,$sec\!\cdot\!core$)} & \multicolumn{3}{c}{strong scaling (time steps\,/\,$sec$)} \\
  \cmidrule(rl){2-4}
  \cmidrule(rl){5-7}
  cores & MPI only & OpenMP (4) & OpenMP (8) & MPI only & OpenMP (4) & OpenMP (8) \\
  \midrule
  32     & 1.72690 & 2.06960 & 1.70424 & & & \\
  64     & 1.72024 & 2.05401 & 1.75702 & & & \\
  128    & 1.70207 & 2.04498 & 1.67617 & & & \\
  256    & 1.70720 & 2.00083 & 1.71336 & & & \\
  512    & 1.69481 & 1.98645 & 1.70756 & \tablenum[table-format=2.3]{ 4.083} & \tablenum[table-format=2.3]{ 4.937} & \tablenum[table-format=2.3]{ 4.942} \\
  1024   & 1.70529 & 1.98478 & 1.67012 & \tablenum[table-format=2.3]{ 6.804} & \tablenum[table-format=2.3]{ 7.961} & \tablenum[table-format=2.3]{ 8.641} \\
  2048   & 1.67337 & 1.97916 & 1.67471 & \tablenum[table-format=2.3]{10.464} & \tablenum[table-format=2.3]{12.460} & \tablenum[table-format=2.3]{13.280} \\
  4096   & 1.66438 & 1.97675 & 1.67879 & \tablenum[table-format=2.3]{15.983} & \tablenum[table-format=2.3]{21.644} & \tablenum[table-format=2.3]{23.128} \\
  8192   & 1.65579 & 1.97816 & 1.67131 & \tablenum[table-format=2.3]{20.527} & \tablenum[table-format=2.3]{30.990} & \tablenum[table-format=2.3]{30.129} \\
  16384  & 1.62633 & 1.95565 & 1.65981 & \tablenum[table-format=2.3]{26.633} & \tablenum[table-format=2.3]{45.913} & \tablenum[table-format=2.3]{46.443} \\
  32768  & 1.62369 & 1.94555 & 1.67588 & \tablenum[table-format=2.3]{34.335} & \tablenum[table-format=2.3]{58.519} & \tablenum[table-format=2.3]{65.057} \\
  65536  & 1.60442 & 1.94437 & 1.66758 & \tablenum[table-format=2.3]{39.391} & \tablenum[table-format=2.3]{70.840} & \tablenum[table-format=2.3]{73.794} \\
  131072 & 1.58830 & 1.93697 & 1.65841 & \tablenum[table-format=2.3]{46.231} & \tablenum[table-format=2.3]{78.146} & \tablenum[table-format=2.3]{81.712} \\
  262144 & 1.58418 & 1.92665 & 1.65170 & \tablenum[table-format=2.3]{51.478} & \tablenum[table-format=2.3]{81.019} & \tablenum[table-format=2.3]{89.843} \\
  458752 & ---     & 1.92446 & 1.64429 &               ---                   & \tablenum[table-format=2.3]{89.376} & \tablenum[table-format=2.3]{98.017} \\
  \bottomrule
  \end{tabular}
\end{table}

\begin{figure}[tbp]
  \centering
  \includegraphics{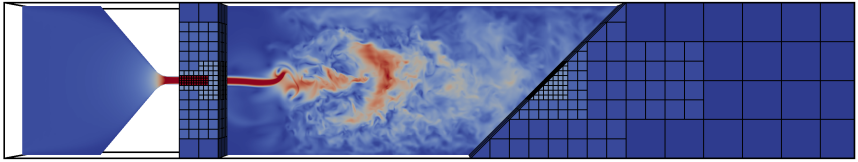}
  \caption{Domain partitioning and flow profile of a preliminary study on the vocal fold and voice generation.
Only the block partitioning is shown.
The simulation uses five levels of refinement.
The highest resolution is required only in the gap between the two folds (blocks highlighted in red).}
  \label{fig:vocal}
\end{figure}

Finally, we turn to an application-oriented example to demonstrate the potential of the presented algorithms.
Figure~\ref{fig:vocal} illustrates a phantom geometry of the vocal fold of a human as it is used to study the voice generation within the human throat~\cite{becker2009}.
Using grid refinement with five different levels of resolution, we can perform direct numerical simulations of flows with a Reynolds number of 1,000.
Here, the domain is partitioned into 25,800 blocks with $16 \times 16 \times 16$ cells per block, resulting in a total of 105,676,800 cells.
With 4,300 processes on SuperMUC, we achieve close to 100 time steps per second.
Due to five levels of refinement, 55.2 times less memory is required and
98.2 times less workload is generated compared to the same simulation with the entire domain refined to the finest level.

\section{Conclusion and outlook}\label{sec:conclusion}

In this article, we have presented the main building blocks that we have implemented into the software framework \Walberla{} 
that enable massively parallel simulations with the LBM on non-uniform grids.
We have introduced the key concepts behind our distributed data structures including a forest of octrees-based domain partitioning into blocks that
provides the foundation for excellent scalability to current supercomputers.
Using these data structures as a basis, we have discussed a distributed memory parallelization approach for the grid refinement technique developed in \cite{Rohde2006}.
Furthermore, we have presented an asynchronous communication scheme that requires only minimal communication and
a level-wise load balancing strategy that perfectly fits the structure of the algorithms involved.
Additionally, we have proposed a method for scaling the two relaxation parameters of the two-relaxation-time collision model across different grid resolutions.

We have verified the correctness of our parallel scheme for two Poiseuille flow scenarios.
Furthermore, we have demonstrated not only near-perfect weak scalability on two current petascale supercomputers but also
an absolute performance of close to a trillion cell updates per second
for a simulation with almost two million concurrent threads and a total number of 886 billion cells.
To our best knowledge, this is the largest computation to date with the LBM on non-uniform grids,
significantly exceeding the data previously published~\cite{Freudiger08,Schoenherr20113730,FLD:FLD2469,Hasert2014784}.
Strong scaling a simulation with several millions of cells, we have reached a performance of less than one millisecond per time step.
A hybrid parallel execution model where multiple OpenMP threads are executed per MPI process
proves to be essential for best performance, especially when strong scaling simulations to massively parallel systems.

Future work will deal with the support of dynamic grid refinement and runtime adaptivity.
To this end, the data structure managing the block partitioning must allow for splitting, merging, and exchanging blocks between different processes at runtime.
As a result, dynamic load balancing strategies must be implemented.
Since our data structure allows to run octree-based as well as general graph-based algorithms (see section~\ref{sec:pconcept:octree}),
dynamic load balancing can be based on space filling curves following, e.g., ideas published in~\cite{Burstedde2011},
or it can be based on diffusive algorithms discussed, e.g., in~\cite{4536237,5395240}.

\hspace{0.5cm}

{\bf Reproducibility.}
All algorithms presented in this article are implemented in the \Walberla{} software framework.
All the benchmarks used in section~\ref{sec:benchmarks} were also added to the framework and are available as part of the software.
Researchers are encouraged to contact the authors of the framework through http://walberla.net in order to get access to the code.

\hspace{0.5cm}

{\bf Acknowledgments.}
The authors would like to thank David Staubach, Ehsan Fattahi, Christian Godenschwager, Regina Ammer, Simon Bogner, and Martin Bauer for valuable discussions.
In addition, we would like to thank two anonymous reviewers, whose remarks helped to significantly improve parts of this article.
We are also grateful to the J\"ulich Supercomputing Center and the Leibniz Rechenzentrum in Munich for providing computational resources.

\bibliographystyle{siam}

\end{document}